\documentclass{emulateapj}
\usepackage{amsmath}
\usepackage{latexsym}
\usepackage{graphicx}

\slugcomment{Submitted for publication in the Astrophysical Journal}

\shorttitle{Collapse of supermassive stars}
\shortauthors{Montero, Janka and M{\"u}ller}

\begin{document}

%%%%%%%%%%%%%%%%%
%%%   TITLE   %%%
%%%%%%%%%%%%%%%%%

\title{Relativistic collapse and explosion of rotating supermassive stars with thermonuclear effects}

\author{Pedro J. Montero\altaffilmark{1},
        Hans-Thomas Janka\altaffilmark{1},
        and
        Ewald M\"{u}ller\altaffilmark{1}
        }
\altaffiltext{1}{Max-Planck-Institut f\"ur Astrophysik,
        Karl-Schwarzschild-Str. 1, D-85748 Garching, Germany;}

\email{montero@mpa-garching.mpg.de}

\begin{abstract}

 We present results of general relativistic simulations of collapsing
 supermassive stars with and without rotation using the two-dimensional general relativistic
 numerical code Nada, which solves the Einstein equations written in
 the BSSN formalism and the general relativistic hydrodynamics
 equations with high resolution shock capturing schemes. These
 numerical simulations use an equation of state which
 includes effects of gas pressure, and in a tabulated form those associated
 with radiation and the electron-positron pairs. We also take into account
 the effect of thermonuclear energy released by hydrogen and helium burning. We find that objects with
 a mass of $\approx 5 \times 10^{5} \rm {M_{\odot}}$  and an initial metallicity
 greater than $Z_{CNO} \approx 0.007$ do explode if non-rotating, while the
 threshold metallicity for an explosion is reduced to $Z_{CNO} \approx 0.001$
 for objects uniformly rotating. The critical initial metallicity for
 a thermonuclear explosion increases for stars with  mass $\approx 10^{6} \rm {M_{\odot}}$.
 For those stars that do not explode we follow the evolution beyond the phase of black hole
 formation. We compute the neutrino energy loss rates due to several processes that may be relevant during the
 gravitational collapse of these objects.  The peak luminosities of
 neutrinos and antineutrinos of all flavors for
 models collapsing to a BH are $L_{\nu}\sim 10^{55} \rm{erg/s}$. The total radiated energy in neutrinos
 varies between $E_{\nu}\sim 10^{56}$ ergs for models collapsing to a
 BH, and $E_{\nu}\sim 10^{45}-10^{46}$ ergs for models exploding.

\end{abstract}

\date{\today}


\section{Introduction}

There is large observational evidence of the presence of  supermassive black
holes (SMBH) in the centres of most nearby galaxies~\citep{Rees98}. The dynamical evidence
related to the orbital motion of stars in the cluster surrounding Sgr
$A^{*}$ indicates the presence of a SMBH with mass
$\approx 4 \times 10^6 \rm {M_{\odot}}$~\citep{Genzel00}.  In addition, the observed
correlation between the central black hole masses and the stellar
velocity dispersion of the bulge of the host galaxies suggests a direct
connection between the formation and evolution of galaxies and SMBH~\citep{Kormendy01}.

The observation of luminous quasars detected at redshifts higher than
6 in the Sloan Digital Sky Survey (SDSS) implies that SMBH with masses $\sim
10^{9} \rm {M_{\odot}}$, which are believed to be the engines of  such
powerful quasars, were formed within the first billion years after the
Big Bang (e.g. Fan 2006 for a recent review). However, it is still an open question how SMBH seeds
form and grow to reach such high masses in such a short amount of
time~\citep{Rees01}.

A number of different routes based on stellar dynamical processes,
hydrodynamical processes or a combination of both have been
suggested (e.g. Volonteri 2010 for a recent review).
One of the theoretical scenarios for SMBH seed formation
is the gravitational collapse of the first generation of stars
(Population III stars) with masses $M \sim 100 \rm {M_{\odot}}$ that are expected to form in halos with
virial temperature $T_{vir} < 10^{4} K$ at $z \sim 20-50$ where
cooling by molecular hydrogen is effective. As a result of the
gravitational collapse of such Pop III stars, very massive BHs would form and then
grow via merger and accretion ~\citep{Haiman01, Yoo04, Alvarez09}.

Another possible scenario proposes that if sufficient primordial gas in massive halos, 
with  mass $\sim 10^{8} M_{\odot}$, is unable to cool below $T_{vir} \gtrsim 10^{4} K$, it may lead to
the formation of a supermassive object \citep{Bromm03,Begelman06}, which would eventually collapse to form a
SMBH. This route assumes that
fragmentation, which depends on efficient cooling, is suppressed, possibly by the
presence of sufficiently strong UV radiation, that prevents the formation of molecular
hydrogen in an environment with metallicity smaller than a given
critical value~\citep{Santoro06,Omukai08}. Furthermore, fragmentation may depend on the turbulence
present within the inflow of gas, and on the mechanism redistributing
its angular momentum~\citep{Begelman09b}. The ``bars-within-bars''
mechanism~\citep{Shlosman89,Begelman06} is a self-regulating
route to redistribute angular momentum and sustain turbulence such
that the inflow of gas can proceed without fragmenting as it collapses
even in a metal-enriched environment.

Depending on the rate and efficiency of the inflowing
mass, there may be different outcomes. A low rate of mass
accumulation would favor the formation of isentropic supermassive
stars (SMS), with mass $\geq 5 \times 10^{4} \rm {M_{\odot}}$, which then would evolve as
equilibrium configurations dominated by radiation
pressure~\citep{Iben63,Hoyle63,Fowler64}. A different outcome could result if the accumulation of gas is fast
enough so that the outer layers of  SMS are not thermally relaxed
during much of their lifetime, thus having an entropy
stratification~\citep{Begelman09}.

 A more exotic mechanism that could eventually lead to a SMS collapsing
 into a SMBH is the formation and evolution of supermassive dark matter stars
 (SDMS) (Spolyar et al 2008). Such stars would be composed primarily
of hydrogen and helium  with only about $0.1\%$ of their mass in the form
of dark matter, however they would shine due to dark matter
annihilation. It has recently been pointed out that SDMSs could reach masses
$\sim 10^{5} \rm {M_{\odot}}$~\citep{Freese10}.  Once SDMSs run out of
their dark matter supply, they experience a contraction phase that
increases their baryon density and temperature, leading to an environment
where nuclear burning may become important for the
subsequent stellar evolution.

If isentropic SMS form, their quasi-stationary
evolution of cooling and contraction will drive the stars to
the onset of a general relativistic gravitational instability leading to their gravitational collapse~\citep{Chandrasekhar64,Fowler64}, and possibly also to the
formation of a SMBH. The first numerical
simulations, within the post-Newtonian approximation, of  \citet{Appenzeller72} concluded that for spherical stars
with masses greater than $10^6 \rm {M_{\odot}}$ thermonuclear reactions have no
major effect on the collapse, while  less massive stars exploded due to
hydrogen burning. Later \citet{Shapiro79} performed the
first relativistic simulations of the collapse of a SMS  in spherical
symmetry. They were able to follow the evolution until the formation of
a BH, although their investigations did not include any microphysics. \citet{Fuller86} revisited the work of \citet{Appenzeller72} and performed simulations of non-rotating SMS in the range of
$10^5$-$10^6 \rm {M_{\odot}}$ with post-Newtonian corrections and detailed
microphysics that took into account an equation of state (EOS) including
electron-positron pairs, and a reaction network describing hydrogen
burning by the CNO cycle and its break-out via the rapid proton
capture (rp)-process. They found that SMS with zero initial metallicity do
not explode, while SMS with masses larger than $10^5 \rm {M_{\odot}}$ and with
metallicity $Z_{CNO} \ge 0.005$ do explode. 

More recently \citet{Linke01} carried out general relativistic
hydrodynamic simulations of the spherically symmetric gravitational
collapse of SMS adopting a spacetime foliation with outgoing null
hypersurfaces to solve the system of Einstein and fluid equations. 
They performed simulations of spherical SMS with masses in the range of $5
\times 10^{5}\rm {M_{\odot}}-10^{9}\rm {M_{\odot}}$ using an EOS that accounts for
contributions from baryonic gases, and in a tabulated form, radiation
and electron-positron pairs. They were able to follow the collapse from the onset of the
instability until the point of BH formation, and showed that an
apparent horizon (AH) enclosing about 25\% of the stellar material was
formed in all cases when simulations stopped.

 \citet{Shibata02} carried
out general relativistic numerical simulations in axisymmetry of the
collapse of uniformly rotating SMS to BHs. They did not take into
account thermonuclear burning, and adopted a $\Gamma$-law
EOS, $P= (\Gamma -1)\rho \epsilon$ with adiabatic
index $\Gamma=4/3$, where $P$ is the pressure, $\rho$ the rest-mass density, and $\epsilon$ the specific
internal energy. Although their simulations stopped before the final
equilibrium was reached, the BH growth was followed until
about 60$\%$ of the mass had been swallowed by the SMBH. They estimated that
about 90$\%$ of the total mass would end up in the final SMBH with a spin
parameter of $J/M^2\sim 0.75$. 
  
The gravitational collapse of differentially rotating SMS in three
dimensions was investigated by \citet{Saijo09}, who focused on
the post-BH evolution, and also on the gravitational wave (GW)
signal resulting from the newly formed SMBH and the surrounding
disk. The GW signal is expected to be emitted in the low frequency LISA band ($10^{-4}-10^{-1}$ Hz).
 
Despite the progress made, the final fate rotating isentropic SMS is still unclear. In particular, it is still an open question for which
initial metallicities hydrogen burning by the
$\beta$-limited hot CNO cycle and its break-out via the $^{15}$O$(\alpha, \gamma)^{19}$Ne
reaction (rp-process) can halt the gravitational collapse of rotating SMS and generate enough thermal energy to lead to an
explosion.  To address this issue, we perform a series of general
relativistic hydrodynamic simulations with a microphysical 
EOS accounting for contributions from radiation,
electron-positron pairs, and baryonic matter,  and taking 
into account the net thermonuclear energy released by the nuclear
 reactions involved in hydrogen burning through the pp-chain, cold and
hot CNO cycles and their break-out by the rp-process, and helium
burning through the 3-$\alpha$ reaction. The numerical simulations were carried out
with the Nada code~\citep{nada},  which solves
the   Einstein   equations  coupled   to   the  general   relativistic
hydrodynamics equations.  

Greek indices run from 0 to 3, Latin
indices from 1 to 3, and we adopt the standard convention for the
summation over repeated indices. Unless otherwise  stated we  use units  in  which $c=G=1$.

%%%%%%%%%%%%%%%%%%%%%%%%%%%%%%%%%%%%%%%%%%%%%%%%%%
\section{Basic equations}
\label{equations}
%%%%%%%%%%%%%%%%%%%%%%%%%%%%%%%%%%%%%%%%%%%%%%%%%%

Next we briefly describe how the system of
Einstein and hydrodynamic equations  are implemented in the
Nada code. We refer to ~\citet{nada} for a more detailed description
of the main equations and thorough testing of the code 
(namely single BH evolutions, shock tubes, evolutions of both spherical and
rotating relativistic stars, gravitational collapse to a BH of a marginally stable spherical
star, and simulations of a system formed by a BH surrounded by a
 self-gravitating torus in equilibrium).

% Spacetime
%-------------------------------------------
\subsection{Formulation of Einstein equations}   
\label{feqs}
%-------------------------------------------

%-------------------------------------------
\subsubsection{BSSN formulation}
%-------------------------------------------

        We follow the 3+1 formulation in which the spacetime is foliated into a set of non-intersecting spacelike hypersurfaces. In
this approach, the line element is written in the following form
\begin{equation}
ds^2 = -(\alpha^{2} -\beta _{i}\beta ^{i}) dt^2 + 
        2 \beta_{i} dx^{i} dt +\gamma_{ij} dx^{i} dx^{j}, 
\end{equation}
where $\alpha$, $\beta^i$ and $\gamma_{ij}$ are the lapse function,  the shift three-vector, and the three-metric, respectively. The latter is defined by 
\begin{equation}
\gamma_{\mu\nu}=g_{\mu\nu}+n_{\mu}n_{\nu},
\end{equation}
where $n^{\mu}$ is a timelike unit-normal vector orthogonal to a spacelike hypersurface.

We make use of the BSSN
formulation~\citep{Nakamura87,Shibata95,Baumgarte99} to solve the
Einstein equations. Initially, a conformal factor $\phi$ is introduced, and the conformally related metric is written as
\begin{equation}
\label{cmetric}
{\tilde{\gamma}}_{ij}=e^{-4\phi}\gamma_{ij}
\end{equation} 
such that the determinant of the conformal metric,
$\tilde{\gamma}_{ij}$, is unity and $\phi=\ln(\gamma)/12$, where $\gamma=\det(\gamma_{ij})$. We also define the conformally related traceless part of the extrinsic curvature $K_{ij}$, 
\begin{equation}
\label{caij}
 {\tilde{A}}_{ij}=e^{-4\phi}A_{ij}=e^{-4\phi}\left(K_{ij}-\frac{1}{3}\gamma_{ij}K\right),
\end{equation}
where $K$ is the trace of the extrinsic curvature. We evolve the
conformal factor defined as $\chi\equiv e^{-4\phi}$ \citep{Campanelli06}, and the auxiliary
variables $\tilde{\Gamma}^{i}$, known as the {\it{conformal connection functions}}, defined as
\begin{equation}
\label{CCF}
\tilde{\Gamma}^{i}\equiv \tilde{\gamma}^{jk}\tilde{\Gamma}^{i}_{\hskip
0.2cm jk}=-\partial_{j}\tilde{\gamma}^{ij},
\end{equation}
where  $\tilde{\Gamma}^{i}_{\hskip 0.2cm jk}$ are the connection
coefficients associated with $\tilde{\gamma}_{ij}$.

During the evolution we also enforce the constraints
${\rm Tr} (\tilde{A}_{ij})=0$  and ${\rm det} (\tilde{\gamma}_{ij})=1$ at
every time step.

 We use the Cartoon method \citep{Alcubierre01b} to impose axisymmetry while using
Cartesian coordinates.

%
%-------------------------------------------
\subsubsection{Gauge choices}
%-------------------------------------------
%
In addition to the BSSN spacetime variables
(${\tilde{\gamma}_{ij}},{\tilde{A}_{ij}},K,\chi,\tilde{\Gamma}^{i}$),
there are two more quantities left undetermined, the  lapse, $\alpha$ and the shift vector $\beta^{i}$. 
We used the so-called {``non-advective 1+log''} slicing \citep{Bona97a}, by dropping   the
advective term in the {``1+log''} slicing condition. In this case, the
slicing condition takes the form
\begin{equation}
\label{1+log1}  
\partial_{t}\alpha = -2 \alpha K. 
\end{equation}
For the shift vector, we choose the ``Gamma-freezing condition'' \citep{Alcubierre03} written as 
\begin{equation}
\label{shift1}  
\partial_{t}\beta^{i} = \frac{3}{4}B^{i},
\end{equation}
\begin{equation}
\label{shift2}  
 \partial_{t}B^{i} = \partial_{t}\tilde{\Gamma}^{i}-\eta B^i,
\end{equation}
where $\eta$ is a constant that acts as a damping term, originally introduced
both to prevent long term drift of the metric functions and to prevent
oscillations of the shift vector.

%-------------------------------------------
\subsection{Formulation of the hydrodynamics equations}    
%-------------------------------------------
 The general relativistic hydrodynamics equations, expressed through the conservation  equations for the stress-energy tensor $T^{\mu\nu}$ and the continuity equation are

\begin{equation}
\label{hydro eqs}
\nabla_\mu T^{\mu\nu} = 0\;,\;\;\;\;\;\;
\nabla_\mu \left(\rho u^{\mu}\right) = 0,
\end{equation}
where $\rho$ is the rest-mass density, $u^{\mu}$ is the fluid
four-velocity and $\nabla$ is the covariant derivative with respect to
the spacetime metric. Following \citet{Shibata03}, the general relativistic hydrodynamics equations are written in a
conservative form in cylindrical coordinates. Since the
Einstein equations are solved only in the $y=0$ plane with Cartesian
coordinates (2D), the hydrodynamic equations are rewritten in Cartesian
coordinates for $y=0$. 
The following definitions for the hydrodynamical variables are used
\begin{equation}
\label{def1}
\rho_{*}\equiv \rho W e^{6\phi},
\end{equation}
\begin{equation}
v^{i}\equiv \frac{u^{i}}{u^{t}}=-\beta^{i}+\alpha\gamma^{ij}\frac{\mbox{\^{u}}_{j}}{hW},
\end{equation}
\begin{equation}
\mbox{\^{u}}_{i}\equiv h u_{i},
\end{equation}
\begin{equation}
\mbox{\^{e}}\equiv
\frac{e^{6\phi}}{\rho_{*}}T_{\mu\nu}n^{\mu}n^{\nu}=hW-\frac{P}{\rho W},
\end{equation}
\begin{equation}
\label{def2}
W\equiv \alpha u^{t},
\end{equation}
where $W$ and $h$ are the Lorentz factor and the specific fluid
enthalpy respectively, and $P$ is the pressure. The conserved variables
 are $\rho_{*}$, $J_{i}=\rho_{*}\mbox{\^{u}}_{i}$, $E_{*}=
   \rho_{*}\mbox{\^{e}}$. We refer to \citet{Shibata03}
for further details.

%%%%%%%%%%%%%%%%%%%%%%%%%%%%%
%\section{Equation of state, neutrino emission and nuclear burning}
\section{Supermassive stars and microphysics}
\label{sec:microphysics}

\subsection{Properties of SMS}
Isentropic SMS are self-gravitating equilibrium configurations of masses in the range
of $10^4 - 10^8 \rm {M_{\odot}}$, which are mainly supported  by radiation pressure, while the pressure of
electron-positron pairs and of the baryon gas are only minor contributions to the
EOS. Such configurations are well described by
Newtonian polytropes with polytropic index $n=3$ (adiabatic index $\Gamma=4/3$). 
The ratio of gas pressure to the total pressure ($\beta$) for
spherical SMS can be written as~\citep{Fowler66}

\begin{equation}
\label{beta_p}
\beta = \frac{P_g}{P_{tot}} \approx \frac {4.3}{\mu}
\left(\frac{M_{\odot}}{M}\right)^{1/2},
\end{equation}
where $\mu$ is the mean molecular weight. Thus $\beta \approx 10^{-2}$
for $M \approx 10^6 \rm {M_{\odot}}$.

Since nuclear burning timescales are too long for $M \gtrsim 10^4 \rm
{M_{\odot}}$, evolution of SMS proceeds on the Kelvin-Helmholtz timescale and
is driven by the loss of energy and entropy by radiation as well as loss of
angular momentum via mass shedding in the case of rotating configurations.

Although corrections due to the nonrelativistic gas of baryons and electrons and general relativistic effects are small,
they cannot be neglected for the evolution. Firstly, gas corrections raise
the adiabatic index slightly above $4/3$

\begin{equation}
\label{ion_gamma}
\Gamma \approx \frac {4}{3}+ \frac{\beta}{6}+0(\beta^2).
\end{equation}

Secondly, general relativistic corrections lead to the existence of a maximum
for the equilibrium mass as a function of the central density. For
spherical SMS this means that for a given mass the star evolves to a critical
density beyond which it is  dynamically unstable against radial
perturbations \citep{Chandrasekhar64}:

\begin{equation}
\rho_{crit}  = 1.994 \times 10^{18}
\left(\frac{0.5}{\mu}\right)^3\left(\frac{M_\odot}{M}\right)^{7/2}
     {\rm g} {\rm cm^{-3}}.
\end{equation}

The onset of the instability also corresponds  to a critical value of the
adiabatic index $\Gamma_{crit}$, i.e. configurations become unstable
when the adiabatic index drops below the critical value

\begin{equation}
\label{Gcri}
\Gamma_{crit} = \frac{4}{3}+1.12\frac{2GM}{Rc^2}.
\end{equation}
This happens when the stabilizing gas contribution to the EOS
does not raise  the adiabatic index above $4/3$ to
compensate for the destabilizing effect of general relativity
expressed by the second term on the righ-hand-side of Eq.~(\ref{Gcri}).

Rotation can stabilize configurations against the radial
instability. The stability of rotating SMS with uniform rotation was
analyzed by \citet{Baumgarte99a,Baumgarte99b}. 
They found that  stars at the onset of the instability have an equatorial radius $R
\approx 640GM/c^2$, a spin parameter $q \equiv cJ/GM^2 \approx
0.97$, and a ratio of rotational kinetic energy  to the gravitational
binding energy of $T/W \approx 0.009$.

\subsection{Equation of State}     

To close the system of hydrodynamic equations (Eq.~\ref{hydro eqs}) we need to define the EOS.  We follow a  treatment which includes
separately the baryon contribution on the one hand, and photons and electron-positron
pairs contributions, in a tabulated form, on the other hand. The baryon
contribution is given by the analytic expressions for the pressure and specific internal energy

\begin{equation}
\label{EOS3.1}
P_{b}=\frac{\mathcal{R}\rho T}{\mu_{b}},
\end{equation}
\begin{equation}
\label{EOS3.2}
\epsilon_{b}=\frac{2}{3}\frac{\mathcal{R}\rho T}{\mu_b},
\end{equation}
where $\mathcal{R}$ the universal gas constant, $T$ the temperature,
$\epsilon_b$ the baryon specific internal energy, and  $\mu_b$ is the mean molecular weight due to ions, which can be
expressed as a function of the mass fractions of hydrogen ($X$), helium ($Y$)
and heavier elements (metals) ($Z_{CNO}$) as 
\begin{equation}
\label{EOS3.2}
\frac{1}{\mu_b}\approx X + \frac{Y}{4}+\frac{Z_{CNO}}{\langle A\rangle},
\end{equation}
where $\langle A\rangle$ is the average atomic mass of the heavy elements. We
assume that the composition of SMS (approximately that of primordial
gas) has a mass fraction of hydrogen $X=0.75-Z_{CNO}$ and  helium $Y=0.25$, where the metallity $Z_{CNO}=1-X-Y$ is an
initial parameter, typically of the order of $Z_{CNO} \sim 10^{-3}$ (see
Table 1 details). Thus, for the initial compositions that we consider the mean molecular
weight of baryons is $\mu_b \approx 1.23$
(i.e. corresponding to a molecular weight for both ions and electrons of $\mu \approx 0.59$).

 Effects associated with photons and the creation of electron-positron
pairs are taken into account employing a tabulated EOS. At temperatures above
$10^{9} \rm{K}$, not all the energy is used to increase the temperature and
pressure, but part of the photon energy is used to create the rest-mass of the electron-positron
pairs. As a result of pair creation, the adiabatic index of the
star decreases, which means that the stability of the star is reduced.

Given the specific internal energy, $\epsilon$ and rest-mass density,
$\rho$, as evolved by the hydrodynamic equations, it is
possible to compute the temperature $T$ by a Newton-Raphson algorithm
that solves the equation $\epsilon^{*}(\rho,T)=\epsilon$ for $T$,

\begin{equation}
\label{EOS3.3}
 T_{n+1} = T_{n} -
\left(\epsilon^{*}(\rho,T_{n})-\epsilon\right) \left.\left(\frac{\partial
  \epsilon^{*}(\rho,T)}{\partial T}\right)\right|_{T_{n}}^{-1},
\end{equation}
where $n$ is the iteration counter.

\subsection{Nuclear burning}
\label{sec:burning}

In order to avoid the small time steps, and
CPU-time demands connected with the solution of a nuclear
reaction network coupled to the hydrodynamic evolution, we apply an
approximate method to take into account the basic effects of
nuclear burning  on the dynamics of the collapsing SMS. We compute the nuclear energy
release rates by hydrogen burning (through the pp-chain, cold and
hot CNO cycles, and their break-out by the rp-process) and helium
burning (through the 3-$\alpha$ reaction) as a function of rest-mass
density, temperature and mass fractions of hydrogen $X$, helium $Y$
and CNO metallicity $Z_{CNO}$. These nuclear energy generation
rates  are added as a source term on the right-hand-side of the
evolution equation for the conserved quantity $E_{*}$. 

The change rates of the energy density 
due to nuclear reactions, in the fluid frame, expressed in units of [$\rm
  {erg}$ $\rm {cm}^{-3}$ $\rm{s}^{-1}$] are given by:

\begin{itemize}

 {\item pp-chain \citep{Clayton83}:}

\begin{align}
\label{pp-chain}
 \left(\frac{\partial e}{\partial t}\right)_{pp} &=
 \rho(2.38\times 10^{6}\rho g_{11}X^{2}T_{6}^{-0.6666} \nonumber \\
  & e^{-33.80/T_{6}^{0.3333}}),
\end{align}
where $T_{6}=T/10^{6}\rm K$, and $g_{11}$ is given by
\begin{align}
\label{g11}
 g_{11} &= 1+0.0123T_6^{0.3333}+0.0109T_6^{0.66666}+\nonumber \\
 & 0.0009T_6.
\end{align}

 {\item 3-$\alpha$ \citep{Wiescher99}:}
\begin{align}
\label{pp-chain}
 \left(\frac{\partial e}{\partial t}\right)_{3\alpha} &=
\rho(5.1\times10^8\rho^2Y^3T_9^{-3} e^{-4.4/T_9}), 
\end{align}
where $T_{9}=T/10^{9}\rm K$.

 {\item Cold-CNO cycle \citep{Shen07}:}

\begin{align}
\label{CCNO}
 \left(\frac{\partial e}{\partial t}\right)_{CCNO} &=
 4.4\times10^{25}\rho^2 X Z_{CNO} \nonumber \\
 & (T_9^{-2/3}e^{-15.231/T_9^{1/3}} + \nonumber \\
 & + 8.3\times10^{-5}T_9^{-3/2}e^{-3.0057/T_9}).  
\end{align}

 {\item Hot-CNO cycle \citep{Wiescher99}:}

\begin{align}
\label{HCNO}
 \left(\frac{\partial e}{\partial t}\right)_{HCNO} &=
4.6\times10^{15}\rho Z_{CNO}.
\end{align}

 {\item rp-process \citep{Wiescher99}:}
\begin{align}
\label{rp}
 \left(\frac{\partial e}{\partial t}\right)_{rp} &=
 \rho(1.77\times 10^{16}\rho Y Z_{CNO} \nonumber \\
 & 29.96 T_9^{-3/2} e^{-5.85/T_9}).
\end{align}

\end{itemize}

Since we follow a single fluid approach, in which we solve only the
hydrodynamics equations~Eq.~(\ref{hydro eqs}) (i.e. we do not solve additional advection equations
for the abundances of hydrogen, helium and metals), the elemental abundances
during the time evolution are fixed. Nevertheless, this assumption
most possibly does not affect significantly the estimate of the threshold metallicity needed
to produce a thermal bounce in collapsing SMS. The average energy
release through the 3-$\alpha$ reaction is about 7.275 MeV for each $^{12}$C
nucleus formed. Since the total energy
due to helium burning for exploding models is $\sim 10^{45}$ ergs (e.g. $9.0 \times 10^{44}$ ergs for
model S1.c); and even considering that this energy is released  mostly in a central region of the SMS containing
$10^4M_{\odot}$ of its rest-mass~\citep{Fuller86}, it is easy to show that the change in the metallicity
is of the order of $10^{-11}$. Therefore, the increase of the
metallicity in models experiencing a thermal bounce is much smaller
than the critical metallicities needed to
trigger the explosions. Similarly, the average change in the mass
fraction of hydrogen due to the cold and hot CNO cycles is expected to be  $\sim 10\%$ for exploding
models. 

\begin{table*}
\begin{center}
\caption{Main properties of the initial models studied. From  left to right the  columns  show: model,
 gravitational mass, initial central rest-mass
  density, $T_{k}/|W|$, angular velocity on the equatorial plane at
  the surface, initial central temperature,
  metallicity, the fate of the star, radial kinetic energy after
  thermal bounce, and total neutrino energy output.}
\label{tab1}
\begin{tabular}{lccccccccccc}
\hline Model & $M$ & $\rho_{c} $  & $T_{k}/|W|$ &
$\Omega $ & $T_{c}$ & Initial metallicity & Fate & $E_{\rm RK}$
& $E_{\nu}$\\
 & $[10^{5}\rm {M_{\odot}}]$ & $[10^{-2}\rm {g/cm^{3}}]$ & & $[10^{-5}
  \rm {rad/s}]$& $[10^7 \rm {K}]$ & $[10^{-3}]$ & &$[10^{56}\rm erg]$ &[erg] \\

\hline
S1.a     & $5$ & $2.4$   & 0
& 0&$5.8$ & $5$ & BH &...& $ 3.4\times 10^{56}$\\
S1.b     & $5 $ & $2.4$ & 0
& 0&$5.8$ & $ 6$ &  BH &  .. & ...\\
S1.c     & $5 $ & $2.4$ & 0
& 0&$5.8$ & $7$ & Explosion
& $ 5.5$ & $9.4 \times 10^{45}$\\
\hline
R1.0     & $5 $ & $40 $ &0.0088 &$2.49$ &$13$ & $ 0 $ & BH & ...&  ... \\
R1.a     & $5 $ & $40 $ &0.0088 &$2.49$ &$13$ & $ 0.5 $ & BH & ...& $ 5.4 \times 10^{56}$ \\
R1.b     & $5 $ & $40 $ &0.0088 &$2.49 $ &$13$ & $0.8 $ & BH & ... & ... \\
R1.c     & $5 $ & $40 $ &0.0088 &$2.49 $ &$13$ & $1$ & Explosion &
$1.0$ & ... \\
R1.d     & $5 $ & $40$ &0.0088 &$2.49 $ &$13$ & $2$ & Explosion & $1.9$ & $ 8.9 \times 10^{45}$ \\
\hline
S2.a     & $10$ & $0.23$ & 0   &0
&$2.6 $ & $30 $ & BH & ...& $ 6.8 \times 10^{56}$\\
S2.b     & $10 $ & $0.23$ & 0   &0
&$2.6 $ & $ 50 $ & Explosion & $ 35$& $ 8.0\times 10^{46}$ \\
\hline 
R2.a     & $10 $ & $12$ & 0.0087 & $1.47$ & $9.7 $ &$0.5$ & 
 BH  & ...& $3.1 \times 10^{56}$\\
R2.b     & $10 $ & $12$ & 0.0087 & $1.47$ & $9.7 $ &$0.8 $ & 
 BH  & ...& ...\\
R2.c     & $10 $ & $12$ & 0.0087 & $1.47 $&$9.7 $ &$1.0 $ &  BH  & ...& ...\\
R2.d     & $10 $ & $12$ & 0.0087 & $1.47$&$9.7 $ &$1.5 $ &  Explosion
&  $1.5$ & $2.1 \times 10^{46}$\\
\hline 
D1     & $5 $ & $6.9 \times 10^4$ & 0.089 & $540$ & $140 $ &$0$ & 
 BH  & ...&...\\
D2     & $6 $ & $1.3 \times 10^5$ & 0.128 & $700$ & $170 $ &$0$ & 

 Stable/BH\footnote{If the contribution of e$^{\pm}$ pairs is not taken into
 account in the EOS (e.g., in the case $\Gamma$-law EOS or an EOS that
includes only the radiation and pressure contributions in an analytic form)
 the model is stable against gravitational collapse. However, if the
 effect of e$^{\pm}$ pairs is considered, the star becomes unstable
 against gravitational collapse due to the reduction of the adiabatic
 index associated with the pair creation.} & ...& ...\\

\end{tabular}
\end{center}
\end{table*}

\subsection{Recovery of the primitive variables}     

After each time iteration the conserved variables
 (i.e.~$\rho_{*},J_{x},J_{y},J_{z},E_{*}$) are
updated and the {\it primitive} hydrodynamical variables (i.e.~$\rho,v^{x},v^{y},v^{z},\epsilon$) have to be
recovered. The recovery is done in such a
way that it allows for the use of a general EOS of the form
$P=P(\rho,\epsilon)$. We calculate a function $f(P^*)=P(\rho^*,\epsilon^*)-P^*$, where
$\rho^*$ and $\epsilon^*$ depend only on the conserved quantities and
the pressure guess $P^*$. The new pressure is computed then iteratively by a
Newton-Raphson method until the desired convergence is achieved.

\subsection{Energy loss by neutrino emission}

The EOS allows us to compute the neutrino losses due to the following
processes, which become most relevant just before BH formation:
\begin{itemize}

{\item  Pair annihilation ($e^{+}+e^{-} \rightarrow \bar{\nu} + \nu$}):
most important process above $10^{9} \rm{K}$. Due to the large mean free path of
neutrinos in the stellar medium at the densities of SMS the energy loss by neutrinos can be significant. For a
$10^{6}\rm {M_{\odot}}$ SMS most of the energy release in the form of
neutrinos originates from this process. The rates are computed using
the fitting formula given by \cite{Itoh96}.

{\item Photo-neutrino emission ($\gamma+e^{\pm} \rightarrow e^{\pm}+ \bar{\nu}
+ \nu$}): dominates at low temperatures $T \lesssim 4\times 10^{8} \rm{K}$ and
 densities $\rho \lesssim 10^{5} \rm{g cm^{-3}}$ \citep{Itoh96}.

 {\item Plasmon decay ($\gamma \rightarrow \bar{\nu}+\nu$}): This is
 the least  relevant process for the conditions encountered by the models we have
 considered because its importance increases at higher densities than
 those present in SMS. The rates are computed using the fitting formula given by \cite{Haft94}. 

\end{itemize}

%%%%%%%%%%%%%%%%%%%%%%%%%%%%%
\section{Computational setup}
\label{sec:numerical-methods}
%%%%%%%%%%%%%%%%%%%%%%%%%%%%%

The  evolution  equations  are integrated  by the  method  of lines,
for which we use an optimal strongly stability-preserving (SSP)
Runge-Kutta algorithm of fourth-order  with
5 stages~\citep{Spiteri02}. We use a second-order slope limiter reconstruction
scheme (MC limiter) to obtain the left and right states of the
primitive variables at each cell interface, and a HLLE approximate
Riemann solver~\citep{Harten83,Einfeldt88} to compute the numerical fluxes in the $x$ and $z$ directions.

Derivative terms in the spacetime evolution equations are represented
by a fourth-order centered finite-difference approximation on a uniform
Cartesian grid except for the advection terms (terms formally like
$\beta^{i}\partial_{i}u$), for which an upwind scheme is used.

The computational domain is defined as $0\leq x \leq L$ and $0 \leq z
\leq L $, where $L$ refers to the location of the outer
boundaries.  We used a cell-centered Cartesian grid to avoid that the
location of the BH singularity coincides with a grid point.

%%%%%%%%%%%%%%%%%%%%%%%%%%%%%%%%%%%%
\subsection{Regridding}
\label{sec:Diag}
%%%%%%%%%%%%%%%%%%%%%%%%%%%%%%%%%%%%

Since  it is not possible to follow the
gravitational collapse of a SMS from the early stages to the phase of
black hole formation with a uniform Cartesian grid (the necessary fine zoning would be computationally too demanding), we adopt a
regridding procedure \citep{Shibata02}. During the initial phase of the collapse we rezone the computational
domain by moving the outer boundary inward, decreasing the grid spacing
while keeping the initial number of  grid points fixed. Initially we
use $N\times N=400 \times 400$ grid points, and place
the outer boundary at $L\approx 1.5 r_{\rm e}$ where $r_{\rm e}$ is
the equatorial radius of the star. Rezoning onto the new grid is done using a polynomial
interpolation. We repeat this procedure 3-4 times until the collapse
timescale in the central region is much shorter than in
the outer parts. At this point, we both decrease the grid spacing and
also increase the number of grid points $N$ in dependence of the lapse
function typically as follows: $N
\times N =800 \times 800$ if $0.8>\alpha>0.6$, $N \times N =1200 \times 1200$ if
$0.6>\alpha>0.4$, and $N \times N =1800 \times 1800$ if
$\alpha<0.4$. This procedure ensures the error in the conservation of the total
rest-mass to be less than $2\%$ on the finest computational domain.

\subsection{Hydro-Excision}
\label{sec:Excision}

To deal with the spacetime singularity from the newly formed BH we use
the method of excising the matter content in a region within the horizon
as proposed by \citet{Hawke05} once an AH is found. This
excision is done only for the hydrodynamical variables, and the
coordinate radius of the excised region is allowed to increase in
time. On the other hand, we do neither use excision nor artificial dissipation terms for the spacetime
evolution, and solely rely on the gauge conditions.

%%%%%%%%%%%%%%%%%%%%%%%%%%%%%%%%%%%%
\subsection{Definitions}
\label{sec:Diag}
%%%%%%%%%%%%%%%%%%%%%%%%%%%%%%%%%%%

Here we define some of the quantities listed in Table 1.  We compute
the total rest-mass $M_{*}$ and  the ADM mass $M$ as
\begin{equation}
\label{Mass}
M_{*}= 4\pi\int_{0}^{L}{x}dx\int_{0}^{L}{\rho_{*}}dz,
\end{equation}
\begin{eqnarray}
\label{MassADM}
M= -2\int_{0}^{L}{x}dx\int_{0}^{L}{}dz\left[-2\pi E
e^{5\phi}+\frac{e^{\phi}}{8}\tilde{R}\right. \nonumber \\ 
\left.-\frac{e^{5\phi}}{8}\left(\tilde{A}_{ij}\tilde{A}^{ij}-\frac{2}{3}K^{2}\right)\right],
\end{eqnarray}
where $E=n_{\mu}n_{\nu}T^{\mu\nu}$ ($n^{\mu}$ being the unit normal to
the hypersurface) and $\tilde{R}$ is the scalar
curvature associated to the conformal metric $\tilde{\gamma}_{ij}$.

The rotational kinetic energy $T_k$ and the gravitational potential
energy $W$ are given by
\begin{equation}
\label{rot_T}
T_k=2\pi\int_{0}^{L}{x^{2}}dx\int_{0}^{L}{\rho_{*}\mbox{\^{u}}_{y}\Omega}dz,
\end{equation}
where $\Omega$ is the angular velocity.
\begin{equation}
\label{W}
W=M-(M_{*}+T_k+E_{int}),
\end{equation}
where the internal energy is computed as

\begin{equation}
\label{Ein}
E_{int}= 4\pi\int_{0}^{L}{x}dx\int_{0}^{L}{\rho_{*}\epsilon}dz.
\end{equation}

In axisymmetry the AH equation becomes a nonlinear
ordinary differential equation for the AH shape function,
$h=h(\theta)$ \citep{Shibata97a,Thornburg07}. We employ  an AH
finder that solves this ODE by a shooting method using 
$\partial_{\theta}h(\theta=0)=0$ and
$\partial_{\theta}h(\theta=\pi/2)=0$ as boundary conditions. We define
the mass of the AH as 
\begin{equation}
\label{Mah}
M_{\rm {AH}}=\sqrt{\frac{\mathcal{A}}{16\pi}},
\end{equation}
where $\mathcal{A}$ is the area of the AH.

\begin{figure}
\includegraphics[angle=0,width=7.5cm]{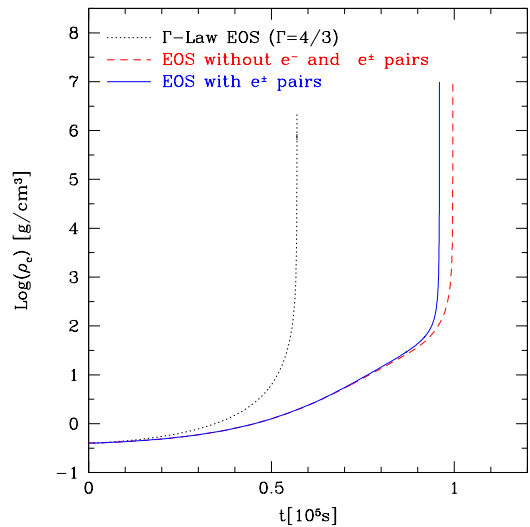}
\caption{ Time evolution of the central rest-mass
density for model R1.0 (a uniformly rotating star with a mass $M = 5 \times
10^{5} \rm {M_{\odot}}$ with zero metallicity) for three different
EOS ($\Gamma$-law and the microphysical EOS with and without
the electron-positron pair creation).}
\label{fig1}
\end{figure}

\begin{figure}
\includegraphics[angle=0,width=7.5cm]{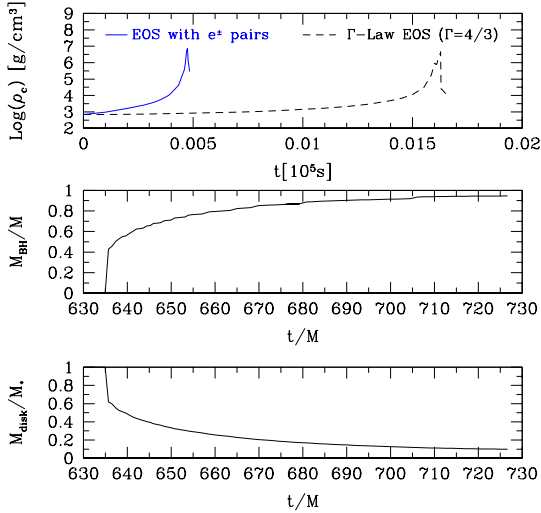}
\caption{The upper panel displays the time evolution of the central rest-mass
density for model D1 (a differentially rotating star with a mass $M
= 5 \times 10^{5} \rm {M_{\odot}}$ and zero metallicity) with a
$\Gamma$-law and the microphysical EOS with electron-positron pair
creation. The middle and lower panels display the AH mass and the
disk mass as a function of time for the collapse simulation with a
$\Gamma$-law EOS.}
\label{fig2}
\end{figure}

%%%%%%%%%%%%%%%%%%%%%%%%%%%%%
\section{Initial models}
\label{sec:initial_models}
The initial SMS are set up as isentropic objects. All models, except
model D2, are chosen such that they are gravitationally unstable,
and therefore their central rest-mass density is slightly larger than the
critical central density required for the onset of the collapse of a
configuration with given mass and entropy. A
list of the different SMS we have considered is provided in Table
1. Models S1 and S2 represent a spherically symmetric, nonrotating SMS with
gravitational mass of $M = 5 \times 10^{5} \rm {M_{\odot}}$ and $M = 1
\times 10^{6} \rm {M_{\odot}}$, respectively, while models R1 and R2 are uniformly
rotating initial models again with masses of  $M = 5 \times 10^{5} \rm {M_{\odot}}$ and $M = 1
\times 10^{6} \rm {M_{\odot}}$, respectively. The rigidly and maximally rotating initial models R1
and R2, and the differentially rotating models D1 and D2 are cconstructed with a polytropic EOS with the Lorene code
(URL http://www.lorene.obspm.fr). We obtain temperatures for our
microphyscial models by inverting the corresponding energy density
with our EOS of Section 3.2. We also introduce a
perturbation to trigger the gravitational collapse by reducing the
pressure overall by $\approx 1.5\%$.

In order to determine the threshold metallicity required to halt the collapse and produce an
explosion we carry out several numerical simulations for each
initial model with different values of the initial metallicity. The
initial metallicities along with the fate of the star are given in
Table 1.

\section{Comparisons with previous studies}
\label{sec:Comparissons}

\subsection{Comparison with 1D calculations}

Axisymmetric calculations without rotation (i.e. models S1 and S2) retain
the spherical symmetry of the initial conditions. There are no
physical phenomena like convective or overturn
instabilities \footnote{In a core-collapse supernova nonradial
 instabilities are triggered either by negative entropy gradients
  caused by the shock deceleration and neutrino heating or by a generic
  instability of the stalled shock (SASI, Blondin and Mezzacapa
  2003). Conditions for both processes are absent in the collapse and
  explosion of SMS.} to produce asphericity, i.e.  we can directly compare our 2D non-rotating models
with those computed in spherical symmetry (1D calculations) by
~\cite{Fuller86} and \citet{Linke01}.

The main differences with respect to the results obtained by
~\cite{Fuller86} are most likely due to two reasons. First, we apply a
 fully general relativistic treatment while they used a post-Newtonian treatment
of gravity, and second, there are differences in the treatment of nuclear burning
(Fuller et al. 1986 solved the relevant nuclear network without the
approximations adopted in our work, see Section 3.3 for further details). Despite
these differences, the results agree fairly well. As discussed
in detail in Section 7.1 the initial metallicities required to
produce an explosion are similar, and  a thermal bounce can be
produced only if sufficient energy is liberated during the phase when the HCNO cycle is active.  

The main difference with respect to the work of ~\cite{Linke01}
resides in the formulation of Einstein's field equations; in
particular, in the foliation of the spacetime (foliation into a set spacelike
hypersurfaces versus a foliation with outgoing null hypersurfaces). In
order to compare with the results of \citet{Linke01}, we computed the redshifted total energy output for
a model having the same rest-mass, doing the time integration until approximately the same
evolutionary stage as in \citet{Linke01}. We find that the total energies released in
neutrinos differ by less than $10\%$ (for mode details see Section 7.3).

\begin{figure}
\includegraphics[angle=0,width=7.5cm]{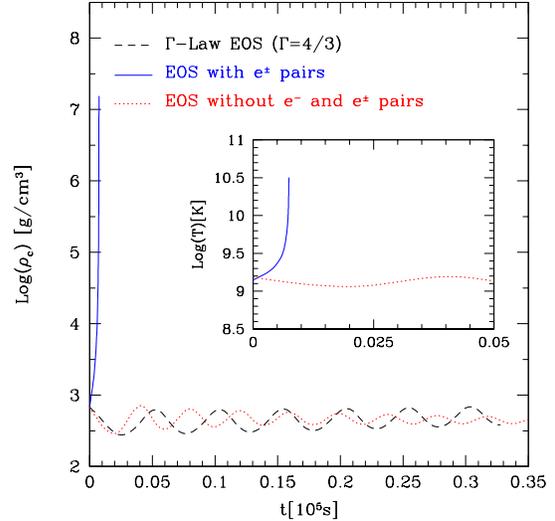}
\caption{ Time evolution of the central rest-mass
density for model D2 for three EOS, which shows that D2
becomes unstable and collapses to a BH only when the microphysical EOS with
electron-positron pairs is used.}
\label{fig3}
\end{figure}

\begin{figure*}
\hskip 1. cm
\includegraphics[angle=0,width=7.5cm]{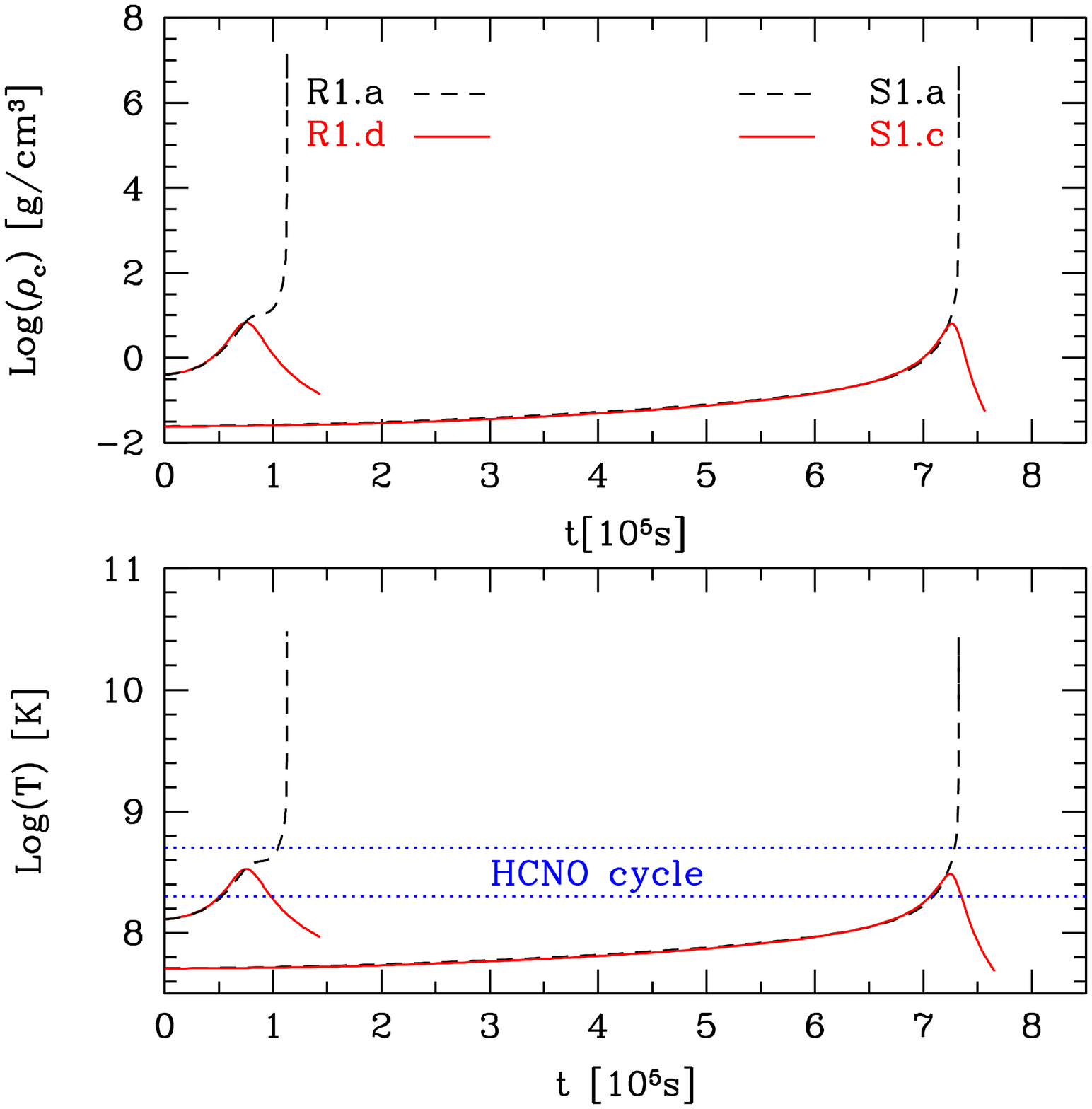}
\hskip 1. cm
\includegraphics[angle=0,width=7.5cm]{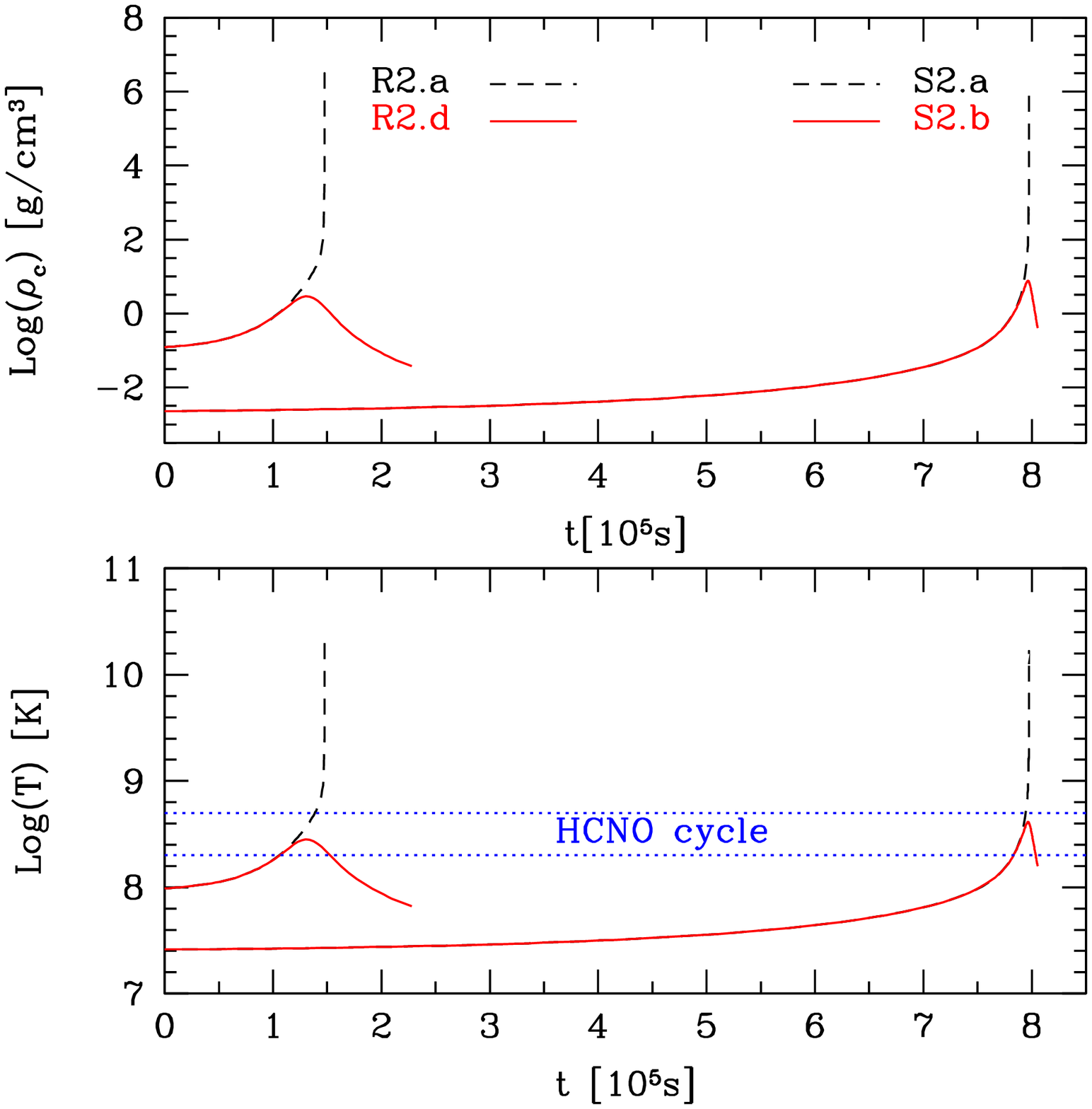}
\caption{Left upper panel shows the time evolution of the central rest-mass
density for models S1 and R1 (i.e., spherical and rotating stars with
mass $M = 5 \times 10^{5} \rm {M_{\odot}}$), and the lower left panel shows the time
evolution of the central temperature. Horizontal dotted lines mark the temperature range in which
nuclear energy is primarily released by the hot CNO cycle. Similarly,
the time evolution of the same quantities for model S2 and R2 (i.e., spherical and rotating stars with
mass $M = 1\times 10^{6} \rm {M_{\odot}}$) are shown in the upper and
lower right panels. As the collapse
proceeds, the central density and temperature rise rapidly, increasing
the nuclear energy generation rate by hydrogen burning. If 
the metallicity is sufficiently high, enough energy can be liberated to produce a thermal bounce. This is the case
for models S1.c, R1.d, S2.b and R2.d shown here.}
\label{fig4}
\end{figure*}

\subsection{$\Gamma$-law vs. microphysical EOS in uniformly rotating SMS}

Previous simulations of SMS collapse to BH in general relativity have been
performed with a $\Gamma$-law EOS with $\Gamma = 4/3$ (with the
only exception being the work of Linke et al. 2001). In order to elucidate
the influence of the EOS on the dynamics of collapsing
SMS, we performed three simulations of the same initial model
(model R1.0, a marginally unstable uniformly rotating SMS with zero initial metallicity) without
nuclear burning effects, and  with three EOS: a $\Gamma$-law
EOS with $\Gamma = 4/3$ (i.e. a similar set-up as in
Shibata \& Shapiro 2002) and the microphysical EOS with and without
including electrons and the e$^{\pm}$ pairs, i.e. for the last EOS
case we consider Eqs. (19) and (20) for the baryons plus
$\epsilon_{\gamma} = aT^{4}$ and
$P_{\gamma}=\frac{1}{3}\epsilon_{\gamma}$ for the photons (where $a$
is the radiation density constant). We denote
the $\Gamma$-EOS as EOS-0, the full microphysical EOS, our canonical
one for the studies of this work as EOS-1, and the reduced
microphysical case as EOS-2.

In Figure~\ref{fig1} we show with a dotted line the time evolution of the central density of model R1.0 with
EOS-0, and with a dashed  (solid) line the time evolution of the
central density with EOS-2 (EOS-1). The first thing to note is  that the collapse
timescale obtained with the $\Gamma$-law EOS is shorter than that
obtained with the microphysical EOS (both EOS-1 and EOS-2), because the ion
pressure contribution to the EOS raises the adiabatic index above $4/3$
(see Eq.~\ref{ion_gamma}). This increase in the adiabatic
index helps to  stabilize  the star against the gravitational
instability, and therefore delays the collapse.

On the other hand, the effect of pair creation  reduces the adiabatic
index below $4/3$ at $T \gtrsim 10^{9} $K. This explains  the
differences  between the solid and dashed lines in Figure~\ref{fig1}
 at central densities $\rho_{c} \gtrsim  10$ g/cm$^{3}$, which correspond to central
temperatures $T_{c} \gtrsim 10^{9} $K. Once the collapse
enters this regime, pair creation becomes relevant enough to
reduce the adiabatic index below $4/3$, which  destabilizes the
collapsing star. Compared to previous works (Shibata \& Shapiro 2002)
the use of a microphysical EOS instead of a $\Gamma$-law EOS delays
the collapse (mostly due to the baryons while e$^{\pm}$ destabilize)
of an initially gravitationally unstable configuration.

\subsection{$\Gamma$-law vs. microphysical EOS in differentially
  rotating SMS}

We also performed 2D axisymmetric simulations of
differentially rotating SMS. First, we investigated the
influence of the EOS on {\em gravitationally unstable stars} using model D1
as a reference. This model corresponds, within the accuracy to
which the initial conditions can be reproduced, to
a differentially rotating unstable SMS discussed by
~\citet{Saijo09} (i.e. their model I).  Results are displayed in
Figure~\ref{fig2}. In the upper panel of this figure, we show the time evolution of the
central density for model D1 with EOS-0 (dashed line), and with the microphysical EOS-1 (solid line). Opposite to the behavior in the case
of the uniformly rotating SMS R1.0, the collapse
timescale is longer with the $\Gamma$-law than with the microphysical EOS. The reason for this difference is that 
the initial central temperature ( $T_{c} \approx 1.4 \times 10^{9} $K)
of model D1 is an order of magnitude higher than the initial central temperature in R1.0. Therefore, electron-positron pair creation reduces the
stability of the star (by reducing $\Gamma$) already during the initial
stages of the collapse.  This behavior is expected to be present
also in 3D, and since the collapse timescale
is reduced when using the microphysical EOS, nonaxisymmetric
instablilities would have even less time to grow before the formation
of a BH. It reinforces the conclusions
of~\citet{Saijo09}, who showed that the three
dimensional collapse of rotating stars proceeds in an approximately
axisymmetric manner. 

The lower two panels of Figure ~\ref{fig2} display the growth of the
 AH mass and the disk mass (defined as the rest-mass
outside the AH of the newly formed BH) as a function of
time (in units of the gravitational mass for comparison with Figures 9 and 10 of Saijo \& Hawke
2009), respectively. The values of both quantities at the end of the simulation agree, within
a $5\%$ difference, with those obtained 
in 3D by~\citet{Saijo09}. We note that there is also good agreement (less than $5\%$
difference) regarding the time at which an AH is first
detected. These observed small differences are likely due to differences in the initial
models and numerical techniques rather than to the influence
of nonaxisymmtric effects. This suggests that our collapse simulation with the same treatment of
physics yields good agreement with the 3D simulations of ~\citet{Saijo09}

We also investigated the influence of electron-positron pair creation on the
evolution of {\em gravitationally stable differentially rotating} SMS using model D2 which
is similar to the stable differentially rotating model III
of \citet{Saijo09}. We performed three simulations of
model D2 varying the EOS. In Figure ~\ref{fig3} we show the
central rest-mass density as a function of time for a $\Gamma$-law
(dashed line), and for the microphysical EOS-1 (solid line) and
EOS-2 (dotted line). In agreement with
the results obtained by \citet{Saijo09} we find that model D2
represents a stable differentially rotating SMS when a  $\Gamma$-law
is used. A  persistent series of oscillations is triggered by the
initial perturbation in the pressure. This is also the case with the microphysical EOS-2 without the inclusion of
electrons and the e${^\pm}$ pairs. However, the time evolution of D2 is completely
different when e${^\pm}$ pairs are taken into account. The
influence of pairs is large enough to destabilize model D2 against gravitational collapse. We note that unlike all other SMS considered in
this paper, which are $\Gamma=4/3$ models initially unstable to
gravitational collapse, model D2 is an initially stable $\Gamma=4/3$ model which becomes gravitationally unstable only by the
creation of electron-positron pairs  at high
temperatures. Hence, using a microphysical EOS with electron-positron pairs is crucial to determine
the stability of differentially rotating SMS.

We note that the central temperature of the initial models D1 and D2 is of the
order of $\approx 10^{9} $K. At this temperature, the main source of
thermonuclear energy is hydrogen burning via the rp-process. It is
however expected that such SMS would previously experience a phase
of hydrogen burning via the cold and hot CNO cycles which would
significantly affect the evolution of the models such that
configurations with high $T_{c}$ as in models D1 and D2 might never be
reached. Therefore, models D1 and D2 are  not
particularly well suited to investigate the existence of a thermal
bounce during collapse (see Section 7). Exploring in detail the parameter space
for the stability of differentially rotating SMS with the
microphysical EOS, and the existence of
a thermal bounce during the collapse phase depending on the initial
stellar metallicity, is  a major task on its own, which is beyond the scope of this
paper. For these reasons we do not consider differentially
rotating SMS in this work.

%%%%%%%%%%%%%%%%%%%%%%%%%%%%%
\section{Results}
\label{sec:Results}

\subsection{Collapse to BH vs. Thermonuclear explosion}
First we consider a gravitationally unstable spherically symmetric SMS
with a gravitational mass of $M = 5 \times 10^{5} \rm {M_{\odot}}$ (S1.a, S1.b and S1.c), which corresponds to a model
extensively discussed in ~\cite{Fuller86}, and therefore allows for a
comparison with the results presented here.~\cite{Fuller86} found
that unstable spherical SMS with $M = 5 \times 10^{5} \rm {M_{\odot}}$ and
an initial metallicity $Z_{CNO}=2\times 10^{-3}$ collapse to a BH while
models with an initial metallicity $Z_{CNO}=5\times 10^{-3}$ explode due
to the nuclear energy released by the hot CNO burning. They also found
that the central density and temperature at thermal bounce (where the
collapse is reversed to an explosion) are $\rho_{c,b}=3.16$
g/cm$^{3}$ and $T_{c,b}=2.6 \times 10^{8}$ K, respectively.

 The left panels in Figure~\ref{fig4} show the time evolution of the central
rest-mass density (upper panel) and central temperature (lower panel) 
for models S1.a, S1.c, R1.a and R1.d, i.e., non-rotating and rotating 
models with a mass of $M = 5 \times 10^{5}
\rm {\rm {M_{\odot}}}$. In particular, the solid lines represent the
time evolution of the central density and temperature for model S1.c
($Z_{CNO}=7\times 10^{-3}$) and R1.d ($Z_{CNO}=5\times 10^{-4}$). As the collapse
proceeds, the central density and temperature rise rapidly, which
increases the nuclear energy generation rate by hydrogen
burning. Since the metallicity is sufficiently high, enough energy can be liberated to
increase the pressure and to produce a thermal bounce. This is the case
for model S1.c. In Figure~\ref{fig4} we show that a
thermal bounce occurs (at approximately $t\sim 7 \times 10^5$ s)
entirely due to the hot CNO cycle, which is the main source of
thermonuclear energy at temperatures in the range $2 \times 10^{8}
\rm{K} \leq T \leq 5 \times 10^{8} \rm{K}$. The rest-mass density at
 bounce is $\rho_{c,b}=4.8$ g/cm$^{3}$ and the temperature
$T_{c,b}=3.05 \times 10^{8} \rm{K}$. These values, as well as the
threshold metallicity needed to trigger a thermonuclear explosion ($Z_{CNO}=7\times 10^{-3}$), are
higher than those found by ~\cite{Fuller86} (who found that a spherical
nonrotating model with the same rest-mass would explode, if the initial
metallicity was $Z_{CNO}=5\times 10^{-3}$).

On the other hand, dashed lines show the time
evolution of the central density and temperature for model S1.a
($Z_{CNO}=5\times 10^{-3}$). In this case, as well as for model S1.b, the collapse is not halted
by the energy release and  continues until an AH is
found, indicating the formation of a BH.

\begin{figure}
\includegraphics[angle=0,width=8.0cm]{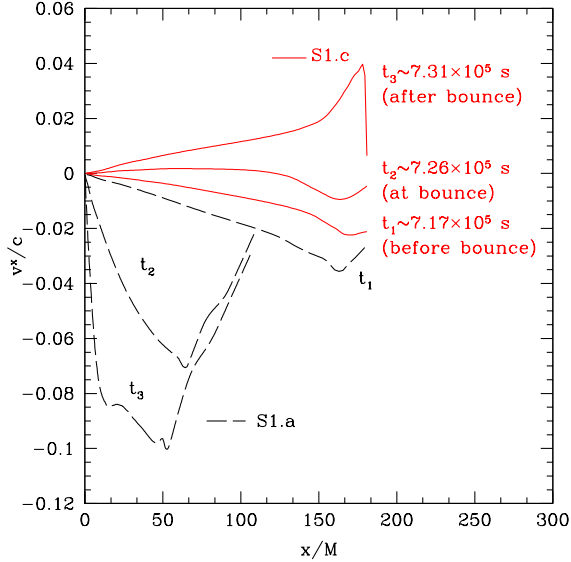}
\caption{Profiles of the $x$-component of the three-velocity $v^{x}$
  along the $x$-axis (in the equatorial plane) for the
  nonrotating spherical stars S1.a (dashed lines) and S1.c (solid
  lines) at three different time
  slices near the time at which  model S1.c experiences a thermal
  bounce. Velocity profiles of model S1.c are displayed up to the
  radius where a shock, that expands into the low density outer layers
  of the SMS, forms.}
\label{fig5}
\end{figure}

We note that the radial velocity profiles change continuously 
  near the time where the collapse is reversed to an explosion due to
  the nuclear energy released by the hot CNO burning, and an
  expanding shock forms only near the surface of the star at a
  radius $R\approx 1.365 \times 10^{13} {\rm cm}$ (i.e. $R/M \approx
  180$) where the rest-mass density is $\approx 3.5 \times 10^{-6}
  {\rm g cm^{-3}}$. We show in Figure~\ref{fig5} the profiles of the $x$-component of the three-velocity $v^{x}$
  along the $x$-axis (in the equatorial plane) for the
  nonrotating spherical stars S1.a (dashed lines) and S1.c (solid lines) at three different time
  slices near the time at which  model S1.c experiences a thermal
  bounce. Velocity profiles of model S1.c are displayed up to the
  radius where a shock forms at $t \approx 7.31 \times
  10^{5} {\rm s}$ and begins to expand into the low density outer layers
  of the SMS.

The evolutionary tracks for the central density and temperature of
the rotating models R1.a and R1.d are also
shown in Figure~\ref{fig4}. A dashed line corresponds 
 to model R1.a, with an initial metallicity $Z_{CNO}=5\times 10^{-4}$, which
collapses to a BH. A solid line denotes  model R1.d with $Z_{CNO}=2\times 10^{-3}$,  which 
explodes due to the energy released by the hot CNO cycle. We find
that Model R1.c with a lower metallicity of $Z_{CNO}=1\times 10^{-3}$
also explodes when the central temperature is the range dominated by
the hot CNO cycle.

As a result of the kinetic energy stored in the rotation of models R1.c and
R1.d, the critical metallicity needed to trigger an  explosion
decreases significantly relative to the non-rotating case. We observe that rotating models with initial
metallicities up to $Z_{CNO}=8\times 10^{-4}$ do not explode
even via the rp-process, which is dominant at temperatures above $T \approx 5 \times
10^{8}$ K and increases the hydrogen burning rate by $200-300$ times
relative to the hot CNO cycle. We also note that the evolution time
scales of the collapse and bounce phases are reduced because rotating models are
more compact and have a higher initial central density and temperature
than the spherical ones at the onset of the gravitational instability.

 The right panels in Figure~\ref{fig4} show the time evolution of the central
rest-mass density (upper panel) and the central temperature (lower panel) 
for models S2.a, S2.b, R2.a and R2.d, i.e., of models with a mass of $M = 10^{6}
\rm {\rm {M_{\odot}}}$. We find that the critical metallicity for an
explosion in the spherical case is $Z_{CNO}=5\times 10^{-2}$ (model
S2.b), while model S2.a with $Z_{CNO}=3\times 10^{-2}$ collapses to a
BH. We note that the critical metallicity leading to a thermonuclear
explosion is higher than the critical value found by~\cite{Fuller86} (i.e., $Z_{CNO}=1\times 10^{-2}$) for a spherical
SMS with the same mass. The initial metallicity leading to an explosion
in the rotating case (model R2.d) is more than an order of magnitude smaller
than in the spherical case. As for the models with a smaller
gravitational mass, the thermal bounce takes place when the physical
conditions in the central region of the star allow for the release of
energy by hydrogen burning through the hot CNO cycle. Overall, the
dynamics of the more massive models indicates that the critical
initial metallicity required to produce an explosion increases with
the rest-mass of the star.

\begin{figure}
\includegraphics[angle=0,width=8.0cm]{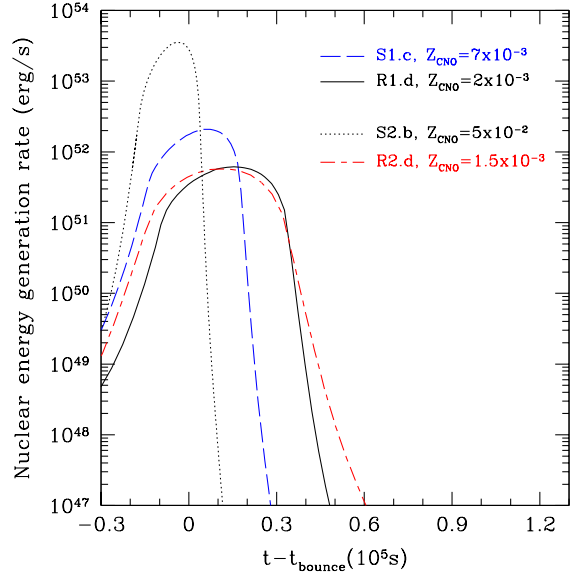}
\caption{Nuclear energy generation rate in erg$/$s for the exploding
  models (S1.c, R1.d, R1.c, S2.b and R2.d) as a function of
time near the bounce. The contribution to the nuclear energy generation is
mainly due to hydrogen burning by the hot CNO cycle. The peak values of the
energy generation rate at  bounce lie between $\approx 10^{51}
[\rm{erg/s}]$ for the rotating models
(R1.d and R2.d), and $\approx 10^{52}-10^{53} [\rm{erg/s}]$ for the
spherical models (S1.c and S2.b).}
\label{fig6}
\end{figure}

Figure~\ref{fig6} shows the total nuclear energy
generation rate in erg$/$s for the exploding models as a function of
time during the late stages of the collapse just before and  after
bounce. The main contribution to the nuclear energy generation is
due to hydrogen burning by the hot CNO cycle. The peak values of the
energy generation rate at bounce lie between several $10^{51}
\rm{erg/s}$ for the rotating models
(R1.d and R2.d), and $\approx 10^{52}-10^{53} \rm{erg/s}$ for the
spherical models (S1.c and S2.b). As expected the maximum nuclear
energy generation rate needed to produce
an explosion is lower in the rotating models. Moreover, as the
explosions are due to the energy release by  hydrogen
burning via the hot CNO cycle, the ejecta would mostly be composed of $^{4}$He.

 As a result of the thermal bounce, the kinetic energy rises
 until most of the energy of the explosion is in the form of kinetic energy.  We list in the
 second but last column of Table 1 the radial kinetic energy after 
 thermal bounce, which ranges between $E_{\rm RK}=1.0 \times 10^{55} \rm {ergs}$ for the
 rotating star R1.c, and  $E_{\rm RK}= 3.5 \times 10^{57} \rm {ergs}$ for
 the spherical star S2.b.

\subsection{Photon luminosity}

\begin{figure*}
\includegraphics[width=6.cm,angle=0]{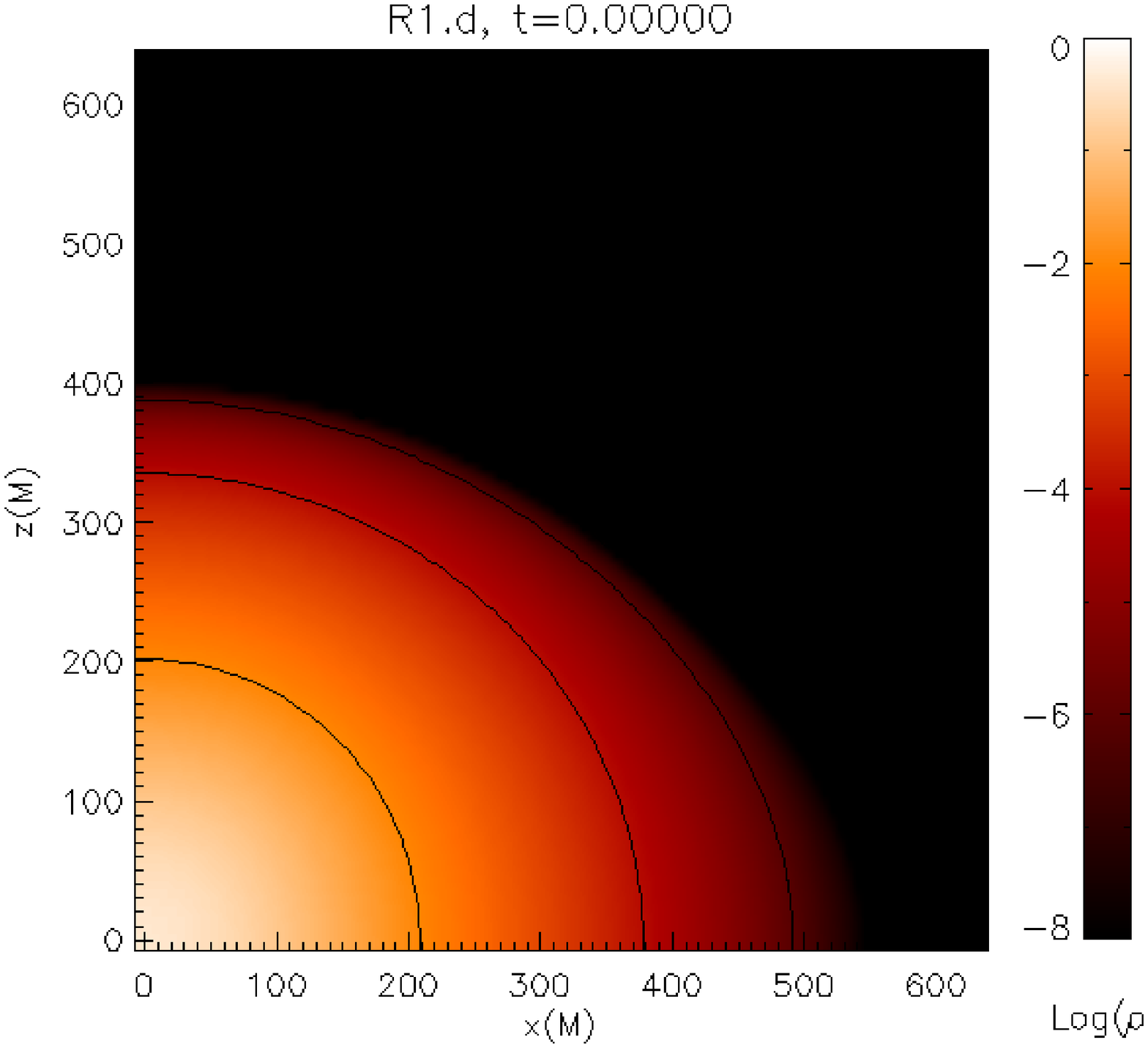}
\includegraphics[width=6.cm,angle=0]{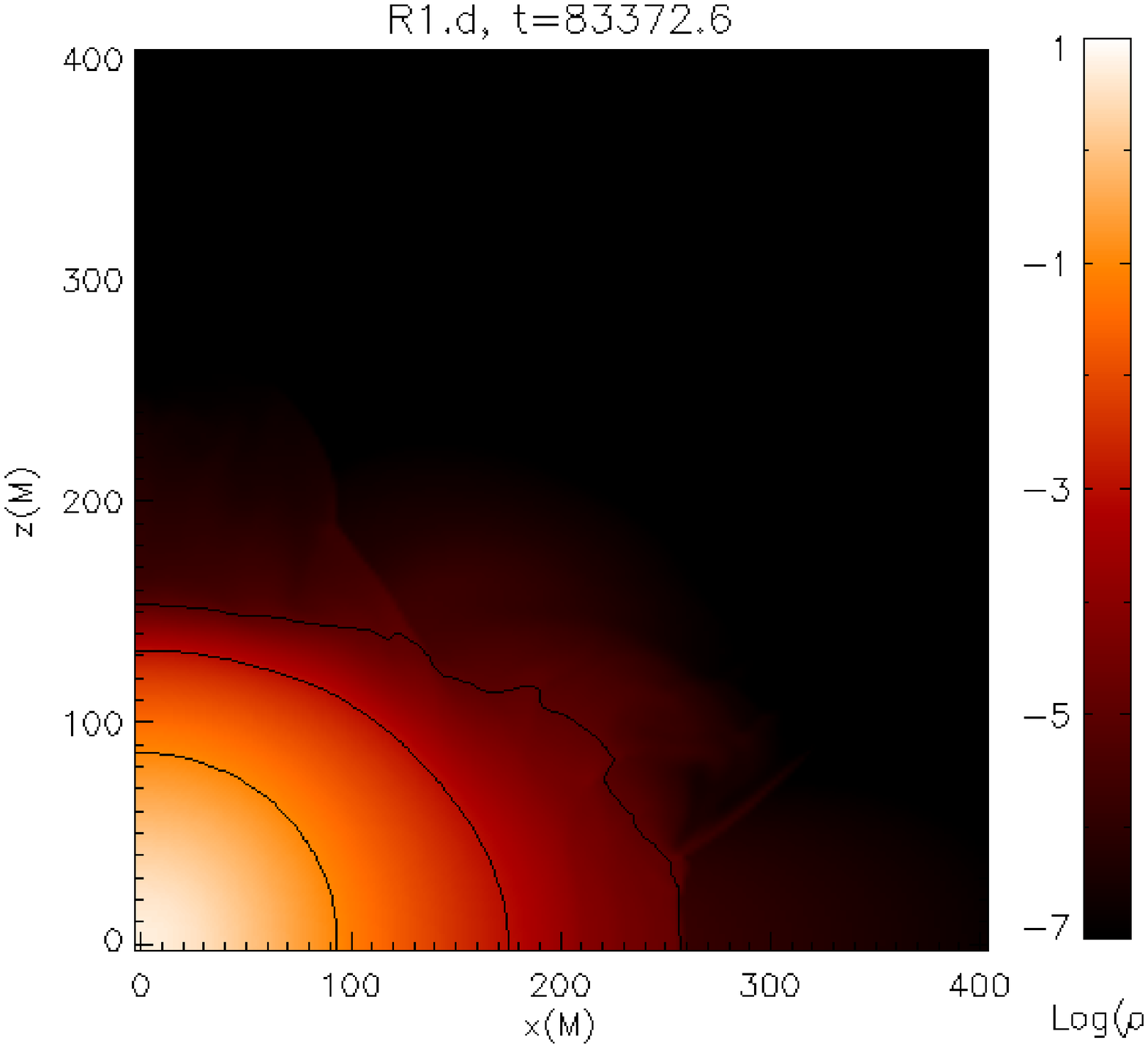}
\includegraphics[width=6.cm,angle=0]{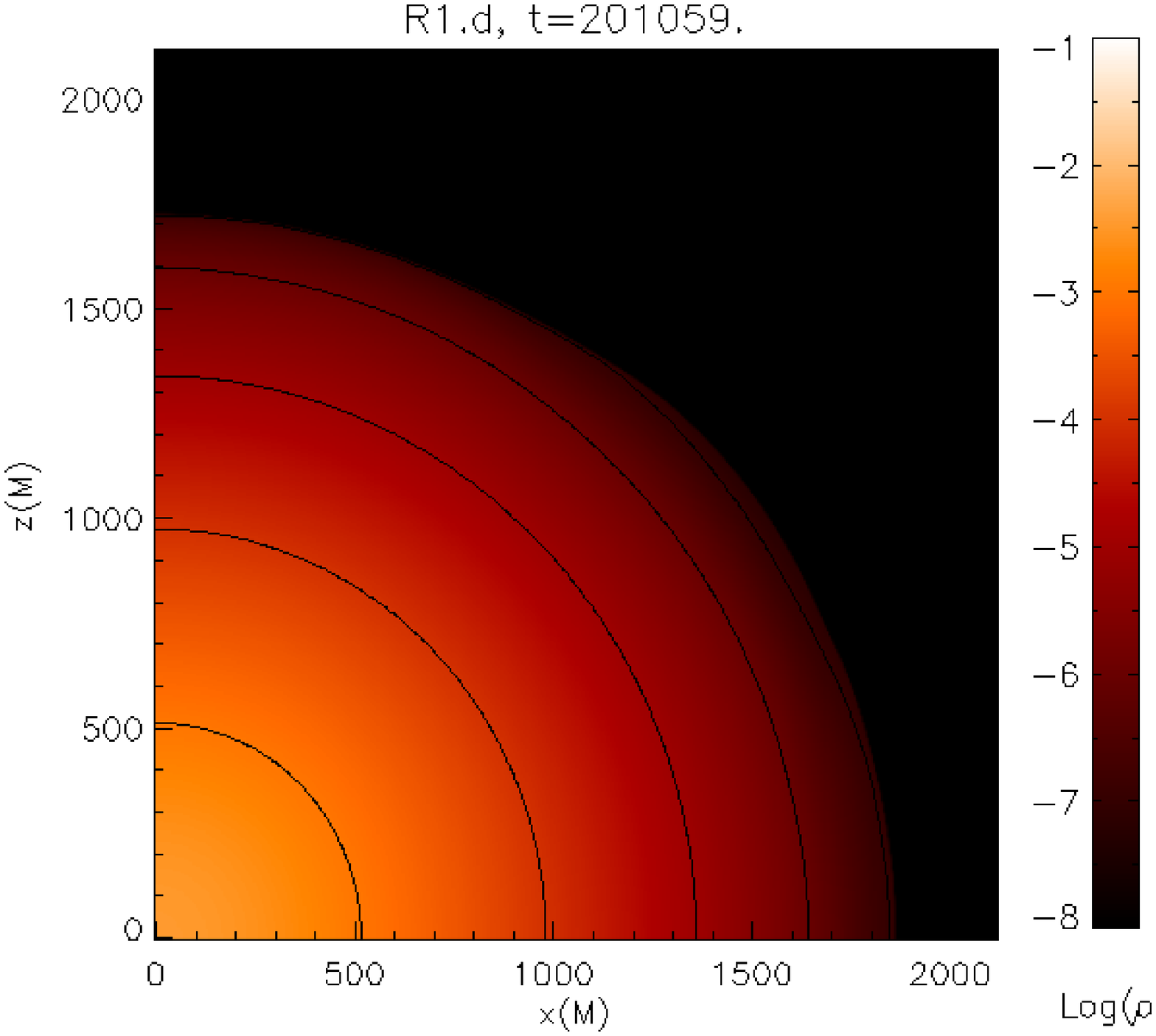}

\caption{ Isodensity contours of the logarithm of the rest-mass density (in g/cm$^{3}$)
  for the rotating model R1.d. The frames have been taken at the initial time (left figure), at
  $t=0.83 \times 10^{5}$ s (central figure) just after a thermal bounce
  takes place, and at $t=2.0 \times 10^5$ s when the radius of the
  expanding matter is roughly 4 times the radius of the star at the
  onset of the collapse.}
\label{fig7}
\end{figure*}

 Due to the lack of resolution at the surface of the star, it becomes
 difficult to compute accurately the photosphere and its effective
 temperature from the criterion that the optical depth is $\tau=2/3$. Therefore,
 in order to estimate the photon luminosity produced in association
 with the thermonuclear explosion, we make use of the fact that within the diffusion
 approximation the radiation flux is given by

\begin{equation}
\label{F_rad}
{F_{\gamma}} =-\frac{c}{3 \kappa_{es}\rho}\nabla{U},
\end{equation}
where $U$ is the energy density of the radiation, and $\kappa_{es}$ is the
opacity due to electron Thompson scattering, which is the main source of
opacity in SMS. The photon luminosity in terms of the temperature gradient and for the
spherically symmetric case can be written as 
\begin{equation}
\label{photon_lum}
{L_{\gamma}} =-\frac{16 \pi a c r^2 T^3}{3
  \kappa_{es}\rho}\frac{\partial{T}}{\partial r},
\end{equation}
where $a$ is the radiation constant, and $c$ the speed of light. As can be seen in the last panel of Figure~\ref{fig7}
the distribution of matter becomes spherically symmetric during the
phase of expansion after the thermonuclear explosion. In this figure
(Fig.~\ref{fig7}) we show the isodensity contours for the rotating
model R1.d. The  frames have been taken at the initial time (left figure), at
  $t=0.83 \times 10^{5}$s (central figure) just after the thermal
bounce (at $t=0.78 \times 10^{5}$s), and at $t=2.0 \times 10^5$ s when the radius of the
  expanding matter is roughly 4 times the radius of the star at the
  onset of the collapse.

\begin{figure}
\hskip 0.5 cm
\includegraphics[angle=0,width=8.0cm]{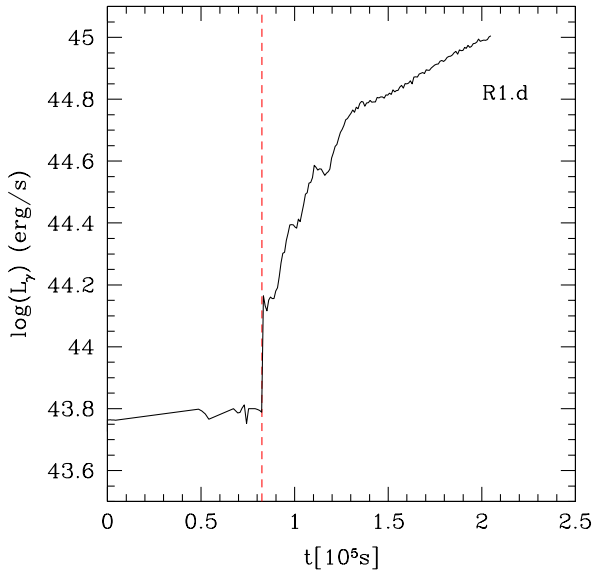}
\caption{Logarithm of the photon luminosity of model R1.d in units of
  $erg/s$ as a function of time. The vertical dashed line indicates
  the time at which the thermal bounce takes place.}
\label{fig8}
\end{figure}

  The photon luminosity computed using Eq.(\ref{photon_lum}) for model R1.d is displayed in Figure~\ref{fig8},
 where we also indicate with a dashed vertical line the time at which
 the thermal bounce takes place. The photon luminosity before the thermal bounce is
 computed at radii inside the star unaffected by the local dynamics of the
 low density outer layers which is caused by the initial pressure
 perturbation and by the interaction between the surface of the SMS
 and the artificial atmosphere. Once the
expanding shock forms near the surface, the photon luminosity is
computed near the surface of the star. The lightcurve shows that, during the initial phase, the luminosity is roughly
equal to the Eddington luminosity $\approx 5\times 10^{43} \rm{erg/s}$
until the thermal bounce. Then, the photon luminosity becomes super-Eddington when the expanding
shock reaches the outer layers of the star and reaches a value of
$L_{\gamma}\approx 1 \times 10^{45} {\rm erg/s}$. This value of the photon
luminosity after the bounce is within a few percent difference with
respect to the photon luminosity \cite{Fuller86} found for a nonrotating SMS of same rest-mass. The photon luminosity
remains super-Eddington during the phase of rapid expansion that follows the thermal
bounce. We compute the photon luminosity until the surface of the star
reaches the outer boundary of the computational domain $\approx 1.0
\times 10^5$ after the bounce. Beyond that
point, the luminosity is expected to decrease, and then rise to a
plateau of $\sim 10^{45} {\rm erg/s}$ due to the
recombination of hydrogen (see Fuller et al. 1986 for a nonrotating
star).

\begin{figure*}
\includegraphics[width=6.cm,angle=0]{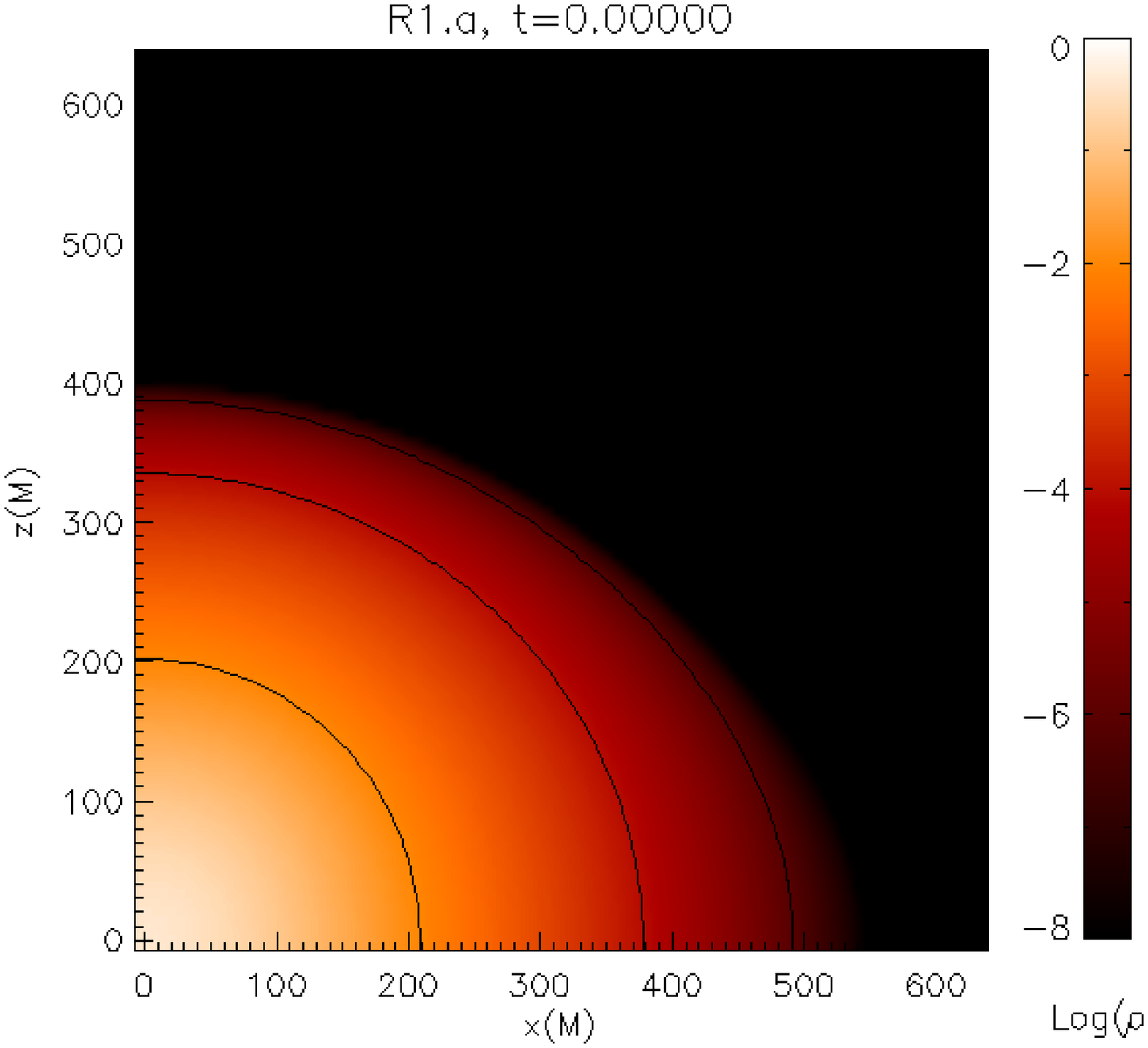}
\includegraphics[width=6.cm,angle=0]{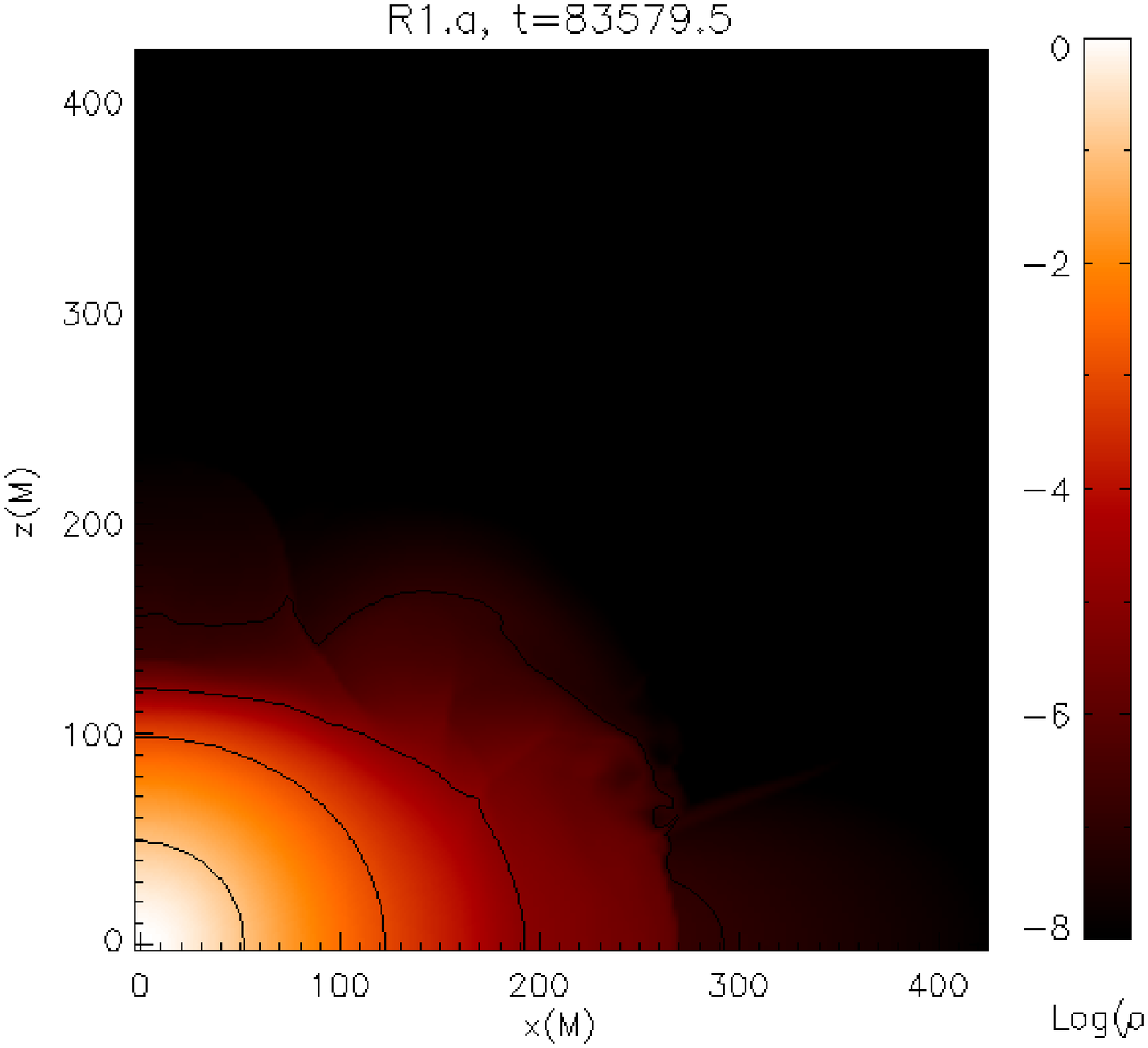}
\includegraphics[width=6.cm,angle=0]{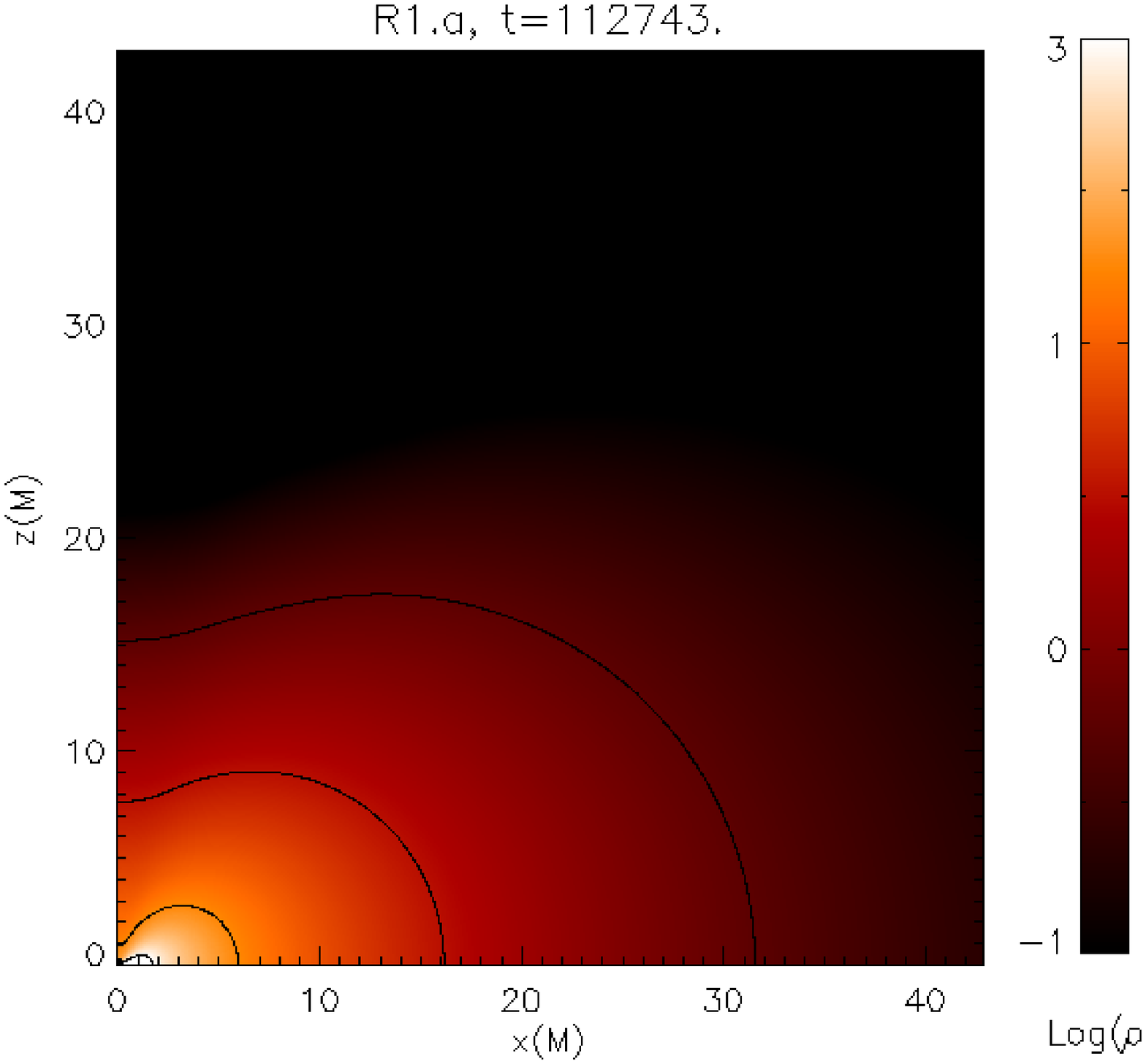}

\caption{Isodensity contours of the logarithm of the rest-mass density
  (in g/cm$^{3}$) for the rotating model R1.a. 
The frames have been taken at the initial time (left panel), at
$t=0.83\times 10^5$ s (central panel), at $t=1.127 \times
10^5$ s, where a BH has already formed and its apparent horizon
encloses a  mass of $50\%$ of the total initial gravitational mass.}
\label{fig9}
\end{figure*}

\subsection{Collapse to BH and neutrino emission}

The outcome of the evolution of models that do not
generate enough nuclear energy during the contraction phase to halt
the collapse is the formation of a BH. The evolutionary tracks for the central density and temperature of
some of these models are also shown in Figure~\ref{fig4}. The central
density typically increases up to $\rho_{c}\sim 10^{7} \rm {g
  cm^{-3}}$ and  the central temperature up to $T_{c}\sim
10^{10}\rm{K}$ just before the formation of an AH. 

Three isodensity contours for the rotating model R1.a collapsing to a BH
are shown in Figure~\ref{fig9}, which display the flattening of the star
as the collapse proceeds. The frames have been taken at the initial time (left panel), at
$t=0.83\times 10^5 \rm s$ (central panel) approximately when model R1.d with
higher metallicity experiences a thermal bounce and, at $t=1.127 \times
10^5 \rm s$, where a BH has already formed and its AH has a mass of
$50\%$ of the total initial mass.

At the temperatures reached during the late stages of the
gravitational collapse (in fact at $T \geq 5 \times 10^{8}\rm{K}$) the
most efficient process for hydrogen burning is the breakout from the hot
CNO cycle via the $^{15}$O$(\alpha, \gamma)^{19}$Ne
reaction. Nevertheless, we find that models  which do not
release enough nuclear energy by the hot CNO cycle to halt their collapse
to a BH, are not able to produce a thermal explosion due to the energy
liberated by the $^{15}$O$(\alpha, \gamma)^{19}$Ne
reaction. We note that above $10^{9} \rm{K}$, not all the liberated energy is used to increase the temperature and
pressure, but is partially used to create the rest-mass of the electron-positron
pairs. As a result of pair creation, the adiabatic index of the
star decreases, which means  the stability of the star is
reduced. Moreover,  due to the presence of e$^{\pm}$ pairs,
neutrino energy losses grow dramatically.

\begin{figure}
\includegraphics[angle=0,width=8.cm]{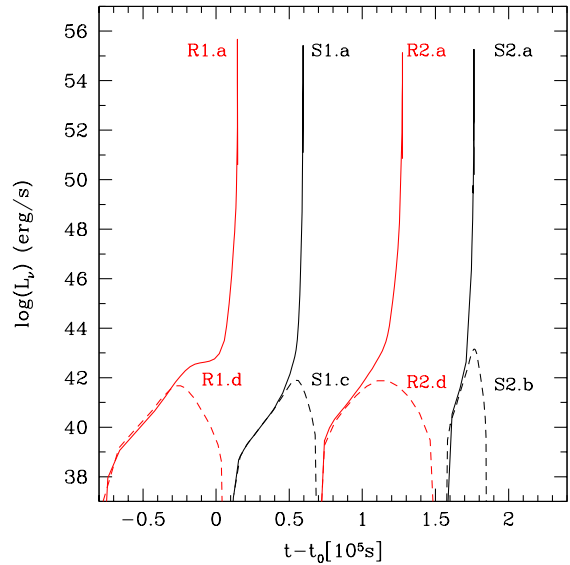}
\caption{Time evolution of the redshifted neutrino luminosities for
  models R1.a, S1.a, R2.a, and S2.a all collapsing to a BH; and for models
  R1.d, S1.c, R2.d, and S2.b experiencing a thermal bounce. The time is measured relative to
  the collapse timescale of each model, R1, S1, R2 and S2, with
  $t_{0}\approx (1,7,0.2,6)$ in units of $10^{5}$s.}
\label{fig10}
\end{figure}

Figure~\ref{fig10} shows (solid lines) the time evolution of the redshifted neutrino luminosities of four
models collapsing to a BH (S1.a, R1.a, S2.a, and R2.a), and (dashed lines) of four models experiencing a
thermal bounce (S1.c, R1.d, S2.b, and R2.d). The change of the slope
of the neutrino luminosities at $\sim 10^{43} \rm {erg/s}$
denotes the transition from  photo-neutrino emission to the
pair annihilation dominated region. The peak luminosities in all form
of neutrino for models collapsing to a BH are $L_{\nu}\sim 10^{55}
\rm{erg/s}$.  Neutrino luminosities can be that important because the
densities in the core prior to BH formation are 
$\rho_{c}\sim 10^{7} \rm {g  cm^{-3}}$, and therefore neutrinos
can escape. The peak neutrino luminosities lie between the
luminosities found by \citet{Linke01} for the collapse of spherical
SMS, and those found by \citet{Woosley86} (who only took into account
the luminosity in the form of electron antineutrino). The maximum luminosity decreases slightly as the rest-mass
of the initial model increases, which was already observed by
~\citet{Linke01}. In addition, we find that the peak of the redshifted
neutrino luminosity does not seem to be very sensitive to the initial
rotation rate of the star. We also note that the luminosity of model R1.a reflects the effects of
hydrogen burning at $L_{\nu}\sim 10^{43} \rm{erg/s}$. 

The total energy output in the form of neutrinos 
is listed in the last column of Table 1 for several models. The total radiated energies vary 
between $E_{\nu}\sim 10^{56}$ ergs for models collapsing to a
BH, and $E_{\nu}\sim 10^{45}-10^{46}$ ergs for exploding models. These
results are in reasonable agreement with previous calculations. For
instance, \citet{Woosley86} obtained that the total energy output in
the form of electron antineutrinos for a spherical SMS with a mass $5
\times 10^{5} {\rm M_{\odot}}$ and zero initial metallicity was $2.6
\times 10^{56} {\rm ergs}$, although their simulations neglected
general relativistic effects which are important to compute accurately
the relativistic redshifts. On the other hand, \citet{Linke01}, by
means of relativistic one-dimensional simulations, found a total
radiated energy in form of neutrinos of about $3 \times 10^{56} {\rm ergs}$ for the
same initial model, and about $1 \times 10^{56} {\rm ergs}$ when
redshifts were taken into account. In order to compare with the results of
\citet{Linke01}, we computed the redshifted total energy output for
Model S1.a, having the same rest-mass, until approximately the same
evolution stage as \citet{Linke01} did (i.e. when the differential
neutrino luminosity $dL_{\nu}/dr
\sim 4 \times 10^{45} {\rm erg/s/cm}$). We find that the total energy released in
neutrinos is $1.1 \times 10^{56} {\rm ergs}$.

The neutrino luminosities for models
experiencing a thermonuclear explosion  (dashed lines in Fig.~\ref{fig10}) peak at much lower values $L_{\nu}\sim 10^{42}-10^{43} \rm{erg/s}$, and  decrease due to the
expansion and disruption of the star after the bounce.

\subsection{Implications for gravitational wave emission}

 The axisymmetric gravitational collapse of rotating SMS with uniform
 rotation is expected  to emit a burst of gravitational
 waves~\citep{Saijo02,Saijo09} with a  frequency within the LISA
 low frequency band ($10^{-4}-10^{-1}\rm {Hz}$). Although through
the simulations presented here we could not investigate the
development of nonaxisymmetric features in our axisymmetric models that could also lead to the
emission of GWs, ~\citet{Saijo09} have shown that the three
dimensional collapse of rotating stars proceeds in an approximately
axisymmetric manner.

In an axisymmetric spacetime, the $\times$-mode vanishes and the $+$-mode of gravitational waves with
$l=2$ computed using the quadrupole formula is written as \citep{Shibata03b}

\begin{equation}
\label{hquad}
h^{quad}_{+} = \frac{\ddot{I}_{xx}(t_{ret})-\ddot{I}_{zz}(t_{ret})}{r}sin^2\theta, 
\end{equation}
where $\ddot{I}_{ij}$ refers to the second time derivative of the
quadrupole moment. The gravitational wave quadrupole amplitude is
$A_2(t)=\ddot{I}_{xx}(t_{ret})-\ddot{I}_{zz}(t_{ret})$. Following
\cite{Shibata03b} we compute the second time derivative of the
quadrupole moment  by finite differencing  the
numerical results for the first time derivative of $I_{ij}$ obtained by

\begin{equation}
\label{Idot}
\dot{I}_{ij}= \int \rho_{*}\left(v^ix^j+x^iv^j\right)d^3x.
\end{equation}

\begin{figure}
\includegraphics[angle=0,width=8.cm]{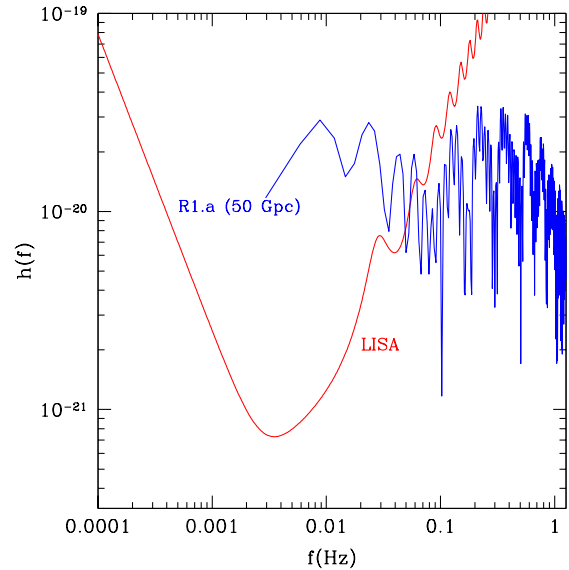}
\caption{Characteristic gravitational wave
strain for model R1.a assuming that the source
is located at a distance of 50 Gpc, together with the design noise spectrum
$h(f)=\sqrt{f S_h(f)}$ for LISA detector.}
\label{fig11}
\end{figure}

We calculate the characteristic gravitational wave strain
\citep{Flanagan98} as

\begin{equation}
\label{h_char}
h_{char}(f) = \sqrt{\frac{2}{\pi}\frac{G}{c^3}\frac{1}{D^2}\frac{dE(f)}{df}},
\end{equation}
where D is the distance of the source, and $dE(f)/df$ the spectral
energy density of the gravitational radiation given by
\begin{equation}
\label{spec_energy}
\frac{dE(f)}{df} = \frac{c^3}{G}\frac{(2\pi f)^2}{16 \pi}\left|\tilde{A}_2(f)\right|^2,
\end{equation}
with
\begin{equation}
\label{fft}
\tilde{A}_2(f) = \int A_2(t)e^{2\pi ift}dt.
\end{equation}
We have calculated the quadrupole gravitational wave emission
 for the rotating model R1.a collapsing to a BH. We plot
in Figure~\ref{fig11} the characteristic gravitational wave
strain (Eq.\ref{h_char}) for this model assuming that the source
is located at a distance of 50 Gpc (i.e., $z \approx 11$) , together with the design noise spectrum
$h(f)=\sqrt{f S_h(f)}$ of the LISA detector \citep{Larson00}. We find
that, in agreement with \cite{Saijo02}, \cite{Saijo09} and \cite{Fryer11}, the burst of
gravitational waves due to the collapse of a rotating SMS could be
detected at a distance of 50 Gpc and at a frequency which approximately
takes the form \citep{Saijo02}

\begin{equation}
\label{fburst}
f_{burst}\sim 3\times 10^{-3}
\left(\frac{10^{6}M_{\odot}}{M}\right)\left(\frac{5M}{R}\right)^{3/2}
{\rm [Hz]},
\end{equation}
where $R/M$ is a characteristic mean radius during black hole
formation (typically set to $R/M=5$).

Furthermore, \citet{Kiuchi11} have
recently investigated, by  means of three-dimensional general relativistic numerical simulations
of equilibrium tori orbiting BHs, the development of the nonaxisymmetric Papaloizou-Pringle
instability (PPI) in  such systems ~\citep{Papaloizou84}, 
and have found that a nonaxisymmetric instability associated with
the $m=1$ mode grows for a wide range of self-gravitating tori orbiting
BHs, leading to the emission of quasiperiodic GWs. In
particular, ~\citet{Kiuchi11} have pointed out that the emission
of quasiperiodic GWs from the torus resulting after the formation of a
SMBH via the collapse of a SMS could be well above the noise sensitivity
curve of LISA for sources located at a distance of 10Gpc.  Such instability
appears for tori whose angular velocity in the equatorial plane
expressed as  $\bar{\Omega}(r)\propto r^{q} $ has $q<q_{kep}$ where
$q_{kep}$ corresponds to the Keplerian limit, i.e. $q = -1.5$ in
Newtonian gravity. 

We find that the torus (defined as the rest-mass outside the AH)
  that forms after the collapse to a BH of
the uniformly rotating model R1.a (when the mass of the AH exceeds
$50\%$ of the ADM mass) does not fulfill the above
condition for the development of the PPI. However, we find that the
torus that forms when the differentially rotating model D1 collapses
to a BH has a distribution of angular momentum such that
$\bar{\Omega}(r)\propto r^{q} $ with $q \approx -1.62$. This suggests
that the torus may be prone to the development of the
nonaxisymmetric PPI, which would lead  to the emission of 
quasiperiodic GWs with peak amplitude $\sim 10^{-18}-10^{-19}$ and frequency $\sim 10^{-3} {\rm Hz}$ 
during an accretion timescale $\sim 10^{5} \rm{s}$.

\subsection{Conclusions}

 We have presented results of general relativistic simulations of collapsing
 supermassive stars using the two-dimensional general relativistic
 numerical code Nada, which solves the Einstein equations written in
 the BSSN formalism and the general relativistic hydrodynamic
 equations with high resolution shock capturing schemes. These
 numerical simulations have used an EOS that
 includes the effects of gas pressure, and tabulated those associated
 with radiation pressure and  electron-positron pairs. We have also taken into account
 the effects of thermonuclear energy release by hydrogen and helium
 burning.  In particular, we have investigated the effects of hydrogen burning by the
 $\beta$-limited hot CNO cycle and its breakout via the $^{15}$O$(\alpha, \gamma)^{19}$Ne
 reaction (rp-process) on the gravitational collapse of nonrotating
 and rotating SMS with non-zero  metallicity.

We have  presented a comparison with previous studies, and
 investigated the influence of the EOS on the collapse. We emphasize that axisymmetric calculations without rotation (i.e. models S1 and S2) retain
 the spherical symmetry of the initial configurations as there are no
 physical phenomena to produce asphericity and numerical artifacts
 associated with the use of Cartesian coordinates are negligibly small. Overall, our collapse
 simulations yield good agreement  with previous works when using the same treatment of
 physics. We have also found that the
 collapse timescale depends on the ion contributions to
 the EOS, and electron-positron pair creation affects the stability of
 SMS. Interestingly, differentially rotating stars that are
 gravitationally stable with a $\Gamma=4/3$ EOS can become unstable against gravitational
 collapse when the calculation is performed with the microphysical EOS
 including  pair creation.

 We have found that objects with a mass of $\approx 5 \times 10^{5}
 \rm {M_{\odot}}$  and an initial metallicity
 greater than $Z_{CNO} \approx 0.007$ explode if non-rotating, while the
 threshold metallicity for an explosion is reduced to $Z_{CNO} \approx 0.001$
 for objects which are uniformly rotating. The critical initial metallicity for
 a thermal explosion increases for stars with a mass of $\approx 10^{6} \rm {M_{\odot}}$. 
 The most important contribution to the nuclear energy generation is
 due to the hot CNO cycle. The peak values of the
 nuclear energy generation rate at bounce range from $\sim 10^{51}
 \rm{erg/s}$ for rotating models
 (R1.d and R2.d), to $\sim 10^{52}-10^{53} \rm{erg/s}$ for spherical 
 models (S1.c and S2.b). After the thermal bounce, the radial kinetic
 energy of the explosion rises  until most of the energy is kinetic, with values
 ranging from $E_{K}\sim 10^{56} \rm {ergs}$ for rotating stars, to up $E_{K}\sim  10^{57} \rm {ergs}$ for
 the spherical star S2.b.  The neutrino luminosities for models
 experiencing a thermal bounce peak at $L_{\nu }\sim 10^{42} \rm{erg/s}$.

 The photon luminosity roughly equal to the
Eddington luminosity during the initial phase of contraction. Then,
after the thermal bounce, the photon luminosity becomes super-Eddington with a value of
about $L_{\gamma}\approx 1 \times 10^{45} \rm {erg/s}$  during the phase of rapid expansion that follows the thermal
bounce. For those stars that do not explode we have followed the evolution beyond the phase of black hole
 formation and computed the neutrino energy loss. The peak neutrino luminosities
 are $L_{\nu}\sim 10^{55} \rm{erg/s}$.

 SMS with  masses less than $\approx 10^{6} \rm {M_{\odot}}$ could
 have formed in massive halos with $T_{vir} \gtrsim 10^{4}
 \rm {K}$. Although the amount of metals that was  
 present in such environments at the time when SMS might have formed
 is unclear, it seems possible that the
 metallicities could have been smaller than the critical metallicities
 required to reverse the gravitational collapse of a SMS into an
 explosion. If so, the final fate of the gravitational collapse of
 rotating SMS would  be the formation of a SMBH and a torus. In a
 follow-up paper, we aim to investigate in detail the
 dynamics of such systems (collapsing of SMS to a BH-torus system) in 3D,
 focusing on the post-BH evolution and the development of nonaxisymmetric
 features that could emit detectable gravitational radiation.

\acknowledgments

We thank B. M{\"u}ller and P. Cerd{\'a}-Dur{\'a}n for useful discussions. Work supported by the Deutsche Forschungsgesellschaft (DFG) through
its Transregional Centers SFB/TR 7 ``Gravitational Wave Astronomy'',
and SFB/TR 27 ``Neutrinos and Beyond'', and the Cluster of Excellence
EXC153 ``Origin and Structure of the Universe''.

\clearpage

\end{document}